\documentclass[12pt]{article}
\usepackage{braket}
\usepackage{multirow}
\usepackage{jheppub}
\usepackage{ifpdf}
\usepackage{booktabs}
\usepackage{array}
 
\usepackage{amssymb}
\usepackage{color}
\usepackage{eufrak}
\usepackage{graphicx,psfrag, subfigure}
\usepackage{amsmath}
\usepackage{upgreek}
\usepackage{bbold}
\usepackage{blkarray}
\usepackage[applemac]{inputenc}
\usepackage{dsfont}
\usepackage{yfonts}
\usepackage{setspace}
\usepackage{soul}
\definecolor{dm}{cmyk}{.20, 0, .30, 0}
\sethlcolor{dm}
\usepackage{mathrsfs}
\numberwithin{equation}{section}
 \usepackage{relsize}
\usepackage{hyperref}
\usepackage{tabularx}

\def\M{M_{{\rm{pl}}}}

\def\be{\begin{equation}}
\def\ee{\end{equation}}
\def\bea{\begin{eqnarray}}
\def\eea{\end{eqnarray}}

\def\be{\begin{equation}}
\def\ee{\end{equation}}
\def\bea{\begin{eqnarray}}
\def\eea{\end{eqnarray}}

\begin{document}

\begin{titlepage}

\setcounter{page}{1} \baselineskip=15.5pt \thispagestyle{empty}

\bigskip\
\begin{center}
{\Large \bf   Inflation Expels Runaways}
\vskip 5pt
\vskip 15pt
\end{center}
\vspace{0.5cm}
\begin{center}
{Thomas C. Bachlechner
}
\end{center}\vspace{0.05cm}

\begin{center}
\vskip 4pt
\textsl{Department of Physics, Columbia University, New York, NY 10027 USA}
\end{center}

{\small  \noindent  \\[0.2cm]
\noindent
We argue that moduli stabilization generically restricts the evolution following transitions between weakly coupled de Sitter vacua and can induce a strong selection bias towards inflationary cosmologies. The energy density of  domain walls between vacua  typically destabilizes K\"ahler moduli and triggers a runaway towards large volume. This decompactification phase can collapse the new de Sitter region unless a minimum amount of inflation occurs after the transition. A stable vacuum transition is guaranteed only if the inflationary expansion generates overlapping past light cones for all observable modes originating from the reheating surface, which leads to an approximately flat and isotropic universe. High scale inflation is vastly favored. Our results point towards a framework for studying parameter fine-tuning and inflationary initial conditions in flux compactifications.
}

\vspace{0.3cm}

\vspace{0.6cm}

\vfil

\begin{flushleft}
\small \today
\end{flushleft}
\end{titlepage}
\tableofcontents
\newpage

\section{Introduction}\label{intro}
On large scales the universe  is extremely well described by an early period of accelerated expansion that evolved into an approximately flat Friedmann-Robertson-Walker cosmology with a small, positive cosmological constant \cite{Perlmutter:1998np,Riess:1998cb,Ade:2015xua,Guth:1980zm,Mukhanov:1981xt,Linde:1981mu,Albrecht:1982wi,Starobinsky:1980te,Guth:1982ec,Hawking:1982cz,Spergel:2003cb,Ade:2015lrj}. 
Despite the marvelous success of the  $\Lambda$CDM model in describing cosmological observations, the associated parameters and initial conditions cannot be explained solely within that model. The small vacuum energy density and the early phase of accelerated expansion both appear rather unnatural. 
We have to revert to a more fundamental description  in order to evaluate how significant the fine-tuning in the effective theory is. 
Constraints imposed by the underlying theory can lead to parameters that would appear surprisingly tuned from a low energy perspective. 
A good example of this mechanism might be the value  of the cosmological constant. If we assume a vast landscape of approximately stable and populated vacua, selection bias  leads to an unnaturally small observed vacuum energy density \cite{Weinberg:1987dv,Bousso:2007kq,DeSimone:2008bq,Bousso:2007er,Clifton:2007bn}. 

In this work we will discuss and employ further assumptions about the fundamental theory. In order to go beyond static  parameters and attempt to constrain cosmological dynamics we consider the following two assumptions in turn:
\begin{enumerate} 
  \item Domain walls between stable de Sitter vacua trigger an instability towards non-positive vacuum energy \cite{Dine:1985he,Cvetic:1994ya,Saffin:1998he,Johnson:2008vn,Aguirre:2009tp,Brown:2011ry}.
  \item The semiclassical mini-superspace approximation applies to  vacuum transition probabilities  \cite{DeWitt:1967yk,Vilenkin:1984wp,Vilenkin:1986cy,Vilenkin:2002ev,BDEMtoappear}.
\end{enumerate}

The first assumption is motivated by  generic instabilities of weakly coupled de Sitter vacua. While flux compactifications exhibit a vacuum structure that may well be able to accommodate a landscape solution to the cosmological constant problem, it is not obvious how the landscape is  populated. 
The low energy theory of flux compactifications contains many zero-energy deformations, referred to as moduli, that need to be stabilized in order to describe well behaved low energy physics. In particular, moduli controlling  perturbative expansions such as the string coupling or the compactification volume are famously difficult to stabilize in a controlled regime.  This observation is known as the Dine-Seiberg problem: weakly coupled vacua are easily susceptible to a runaway instability towards strong or weak coupling \cite{Dine:1985he}. Stable and controlled vacua generically require an accidental cancelation of multiple large terms in the perturbative expansion that renders  the barrier towards  runaway relatively small. 
The large energy density within a domain wall between cosmological vacua can spoil the delicate cancelation and destabilize the moduli. This represents an obstacle to populating the landscape \cite{Johnson:2008vn}. It is conceivable that vacuum transitions at very low energies decouple from potentially unstable moduli, but at high energies the moduli dynamics become increasingly relevant. The relevance of moduli stabilization for cosmology is familiar from models of string inflation. In that context, couplings to moduli at best spoil the  flatness of an inflationary potential and at worst destabilize the entire configuration. Either way, it is crucial to carefully account for the presence of moduli. Even though this observation is well appreciated in inflationary model building, moduli stabilization is often ignored when studying vacuum decay. 

In this work we explore the consequences of moduli instabilities by assuming the absence of stable domain walls between de Sitter vacua in the landscape. Although this means that any vacuum transition triggers an expanding true vacuum bubble containing the runaway  phase, regions of the new de Sitter vacuum are stable at late times if the domain wall remains outside the Hubble horizon. The cosmology is at risk of extinction via domain wall collapse if the horizon expands and allows the runaway domain wall to enter the Hubble sphere. We will find that a vacuum transition is protected against collapse if sufficient inflationary expansion occurs such that all observable modes originating from the end of inflation have overlapping past light cones. This finding coincides with the observational constraint on inflation, but arises independently. In a universe that evolves via inflation and radiation domination to a late time de Sitter phase the lower bound on the number of efolds is set roughly by
\be\label{nebound1}
N_e\gtrsim{1\over 2}\log\left( {H_{\text{Inf}}\over H_{\text{Late}}}\right)\,,
\ee
where $H_{\text{Inf}}$ and $ H_{\text{Late}}$ are the inflationary and late time Hubble parameters, respectively. For high scale inflation late time stability requires inflationary expansion by a factor of roughly $e^{65}$ and  leads to a flat, isotropic universe.  
The lower bound on the required expansion is independent of the bubble nucleation process and applies to Coleman-de Luccia (CDL) transitions  in theories with dynamical gravity \cite{Coleman:1980aw}.

The second assumption concerns the probability of quantum tunneling in a gravitational theory. Even though  our understanding of quantum gravity is tentative, it will be instructive to consider its potential implications. 
We assume that the semiclassical  mini-superspace description is a good guide to compute transition probabilities in spherically symmetric quantum gravity. We employ the canonical formalism to quantize the gravitational theory of spatial geometries, see \cite{DeWitt:1967yk}. 
Following \cite{Vilenkin:1984wp,Vilenkin:1986cy}, we use the WKB ansatz for the  tunneling wave function to approximate transition probabilities through classical potential barriers. In the present work we treat the transition rate as an assumption, while a thorough derivation is presented in \cite{BDEMtoappear}. 
The vacuum transition rate $\Gamma$ within an initial asymptotically flat spacetime is then maximized by the nucleation of a single small region containing the highest inflationary scale,
\be\label{rate1}
\Gamma\sim e^{-\pi /GH_{\text{Inf}}^{2}}\,.
\ee
These transitions to high energy states are most likely to trigger moduli instabilities at the domain walls, consistent with our first assumption.

We will not impose any further assumptions about quantum gravity that are sometimes implicit in the literature, such as conjecturing the absence of bubble nucleation geometries that violate local energy conditions. Even though these configurations cannot arise classically from non-singular initial conditions, quantum fields  violate  local energy conditions \cite{Epstein:1965zza,Bousso:2015wca}. It may be interesting to assume the absence of such tunneling configurations, but this would constitute a third assumption, and given the lack of supporting evidence we elect to avoid it here.

Predictions of physical observables are not possible purely within this simple framework. A prediction would require a detailed understanding of the relevant landscape, vacuum transitions in quantum gravity, the probability measure in eternal inflation and the cosmological dynamics before and after reheating. 
However, by  combining some simple and weak assumptions about fundamental physics we arrive at an educated guess concerning the resulting cosmology: assuming a rich landscape of vacua, the absence of stable domain walls between de Sitter vacua and a vacuum transition rate described by the tunneling wave function, we   expect to find most observers in an isotropic, flat universe with small cosmological constant that experienced an extended period of high scale inflation. Note that all stable vacua can have energy densities well below the scale of inflation. While our speculations shall not be confused with predictions,  an observation of signatures corresponding to high scale inflation would be consistent with our assumptions. 

The organization of this paper is as follows. In \S\ref{domainwallsection} we review the Dine-Seiberg problem of moduli stabilization and the consequences for domain wall stability.  In \S\ref{vacuumtransitionss} we review the gravitational theory of thin, spherically symmetric domain walls. We discuss the classical dynamics and the semiclassical description of vacuum transitions in the Hamiltonian framework. In \S\ref{gravityeffects} we apply the theory of vacuum transitions to the question at hand and demonstrate how gravity can stabilize an otherwise forbidden  transition. Finally, in \S\ref{inflatinsection} we speculate about the cosmological implications for large universes that are stable at late times. We conclude in \S\ref{conclusionss}. Appendices \ref{app1}, \ref{app2} and \ref{app3} provide details and  further explanations that would have distracted from the main text.

\section{Domain Walls and Moduli Stabilization}\label{domainwallsection}
It is instructive to recall some of the basic fields  present in the low energy effective theory corresponding to weakly coupled compactifications of string theory. 
String theory compactified on a Ricci flat manifold without fluxes  contains a potentially large number of moduli that are massless at tree level. The existence of moduli is both good and bad news: on the one hand, all fields need to be stabilized to describe  a realistic cosmology. Light moduli can mediate fifth forces, spoil the inflationary evolution and affect the subsequent expansion of the universe. Even when stabilized at high energy, moduli remain of crucial importance for some low energy phenomena. This is very apparent in  models of string inflation, where a coupling between the volume modulus and the inflaton candidate typically generates a  large slow roll parameter  \cite{McAllister:2005mq}. On the other hand, the existence of hundreds of moduli  can produce a complex  landscape that accommodates the smallness of the observed cosmological constant \cite{Bousso:2000xa,Denef:2006ad,Bachlechner:2015gwa}, and provides for a plethora of potential inflaton candidates \cite{Dimopoulos:2005ac,Arvanitaki:2009fg}. It is therefore imperative to carefully consider the effects of moduli stabilization when studying cosmological models within the string theory framework. 

Some of the most delicate fields are  moduli  controlling perturbative expansions, such as the inverse string coupling $g^{-1}_\text{s}$, or the volume modulus $\rho$. In order for  computable, leading order effects to accurately describe  relevant physics, the effective  potential vanishes in the weak coupling limit. The potential can approach zero  from either side, which induces a runaway instability towards  weak or strong coupling.
 A stable, weakly coupled vacuum can only arise when multiple terms in the perturbative expansion are competitive. Unfortunately, this can signal a breakdown of the  expansion unless one of the terms is accidentally much larger than all consecutive terms. This  phenomenon is known as the Dine-Seiberg problem \cite{Dine:1985he}: most vacua are  expected in strongly coupled regions of moduli space. Despite the challenges in constructing well controlled meta-stable vacua in flux compactifications it is still worthwhile to investigate the cosmology of these models. 

\begin{figure}
  \centering
  \includegraphics[width=1\textwidth]{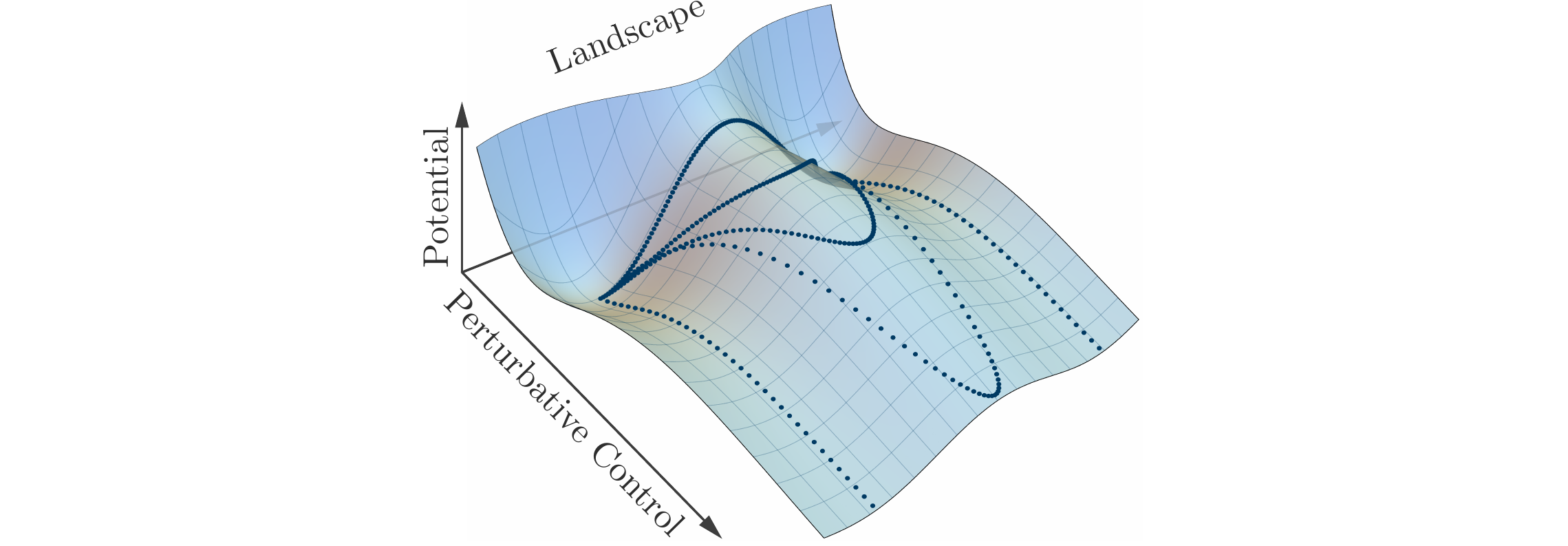}
  \caption{\small Illustration of a potential landscape as a function of a field parametrizing stabilized vacua and a modulus controlling the perturbative expansion. Moduli couplings classically destabilize a domain wall interpolating between vacua of the landscape. \label{kkltwall}}
  \end{figure}
One immediate consequence of the Dine-Seiberg problem is that weakly coupled de Sitter vacua --- if they exist ---  have a relatively small barrier protecting them from a runaway. This observation has severe consequences for transitions between metastable vacua. Consider a landscape with distinct vacua that describe significantly different low energy physics. All moduli are stabilized in each of the minima by some accidental cancelation, but there is no reason to expect a stable domain wall that interpolates between any two vacua: the contributions to the effective potential change significantly along the trajectory in moduli space, exacerbating the Dine-Seiberg problem. Therefore, the presence of a domain wall  between metastable  vacua will generically  de-stabilize some moduli, as illustrated in Figure \ref{kkltwall}. The instability towards a runaway phase leads to the collapse of the  interior de Sitter vacuum and naively poses an obstacle to populating the landscape \cite{Cvetic:1994ya,Saffin:1998he,Johnson:2008vn,Aguirre:2009tp,Brown:2011ry}.

\subsection{Runaway domain walls}
Consider two canonically normalized scalar fields $\phi$ and ${\tau}$ that are decoupled at low energies and interact only via Planck suppressed operators. The field $\phi$ is a proxy for parametrizing a landscape of vacua, while $\tau$  plays the roll of a modulus controlling perturbative expansions, so $\lim_{\tau\rightarrow \infty} V=0$. We are interested in a thin domain wall that does not destabilize the modulus. For simplicity, we  consider a background configuration of $\phi$ containing a  planar domain wall of approximately constant energy density $V_\text{wall}$, and assume a vanishing potential energy in each of the vacua. With only gravitational interactions between the domain wall and the modulus, we expand  the effective potential for the modulus within the domain wall as
\be
V({\tau})\approx V_\text{wall}+{m_{\tau}^2\over 2}{\tau}^2+{V_\text{wall}\tau\over \M}\,.
\ee
To estimate whether the modulus $\tau$ remains in its original vacuum, we denote the  potential barrier of its confining minimum by $V_{\text{barrier}}$. Solving the Klein-Gordon equation for the modulus we find that $\tau$ is pushed beyond its local minimum when the domain wall tension in Planck units exceeds the barrier height\footnote{A slightly different argument arriving at the same conclusion is presented in \cite{Johnson:2008vn}.},
\be\label{criterion}
{\sigma^2\over8\M^2}\gtrsim V_{\text{barrier}}\,,
\ee
where $\sigma=a V_{\text{wall}}$ is the domain wall tension, and $a$ is its spatial thickness. For a domain wall of a scalar field theory we can approximate the domain wall tension as $\sigma\sim \sqrt{V_{\text{wall}}}\Delta\phi$, where $\Delta\phi$ is roughly the canonical length of the trajectory interpolating between the two vacua in field space. 

\begin{figure}
  \centering
  \includegraphics[width=1\textwidth]{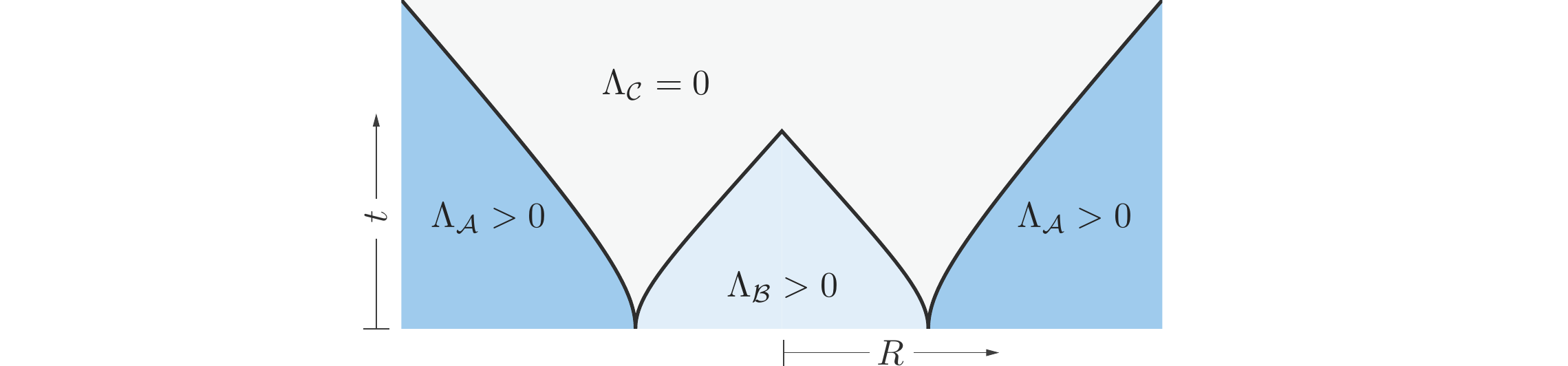}
  \caption{\small Illustration of the evolution of a domain wall between vacua ${\cal A}$ and ${\cal B}$ that is classically unstable towards the runaway phase ${\cal C}$. We neglect any gravitational effects.\label{runawaybubble}}
  \end{figure}

If the vacuum energy differences between metastable minima are set by the typical scale of the potential, $V_{\text{wall}}$, approximate stability requires a low vacuum decay rate, $\log(\Gamma)\sim -\sigma^4/V_{\text{wall}}^3\ll -1$. Combining this bound with (\ref{criterion}) implies a sufficient condition for destabilizing the modulus,
\be
\sqrt{V_{\text{wall}}\over \M^4}\gtrsim {V_{\text{barrier}}\over  V_{\text{wall}}}\,.
\ee
Therefore, if the height of the confining barrier $V_{\text{barrier}}$ is very small compared to the effective potential $V_{\text{wall}}$ away from the vacua, the modulus will be classically destabilized by the presence of a domain wall. This is precisely the situation present in generic weakly coupled vacua of string compactifications and we might not expect stable domain walls interpolating between distinct de Sitter vacua\footnote{There may be accidental cancelations or other effects that reduce the domain wall tension below the naive expectation and leave the modulus stabilized at the domain wall.}. The nucleation of a true vacuum bubble may trigger the runaway instability and could lead to a time-like singularity, as illustrated in Figure \ref{runawaybubble}.
\subsection{Moduli stabilization in IIB string theory}
To see the Dine-Seiberg problem in action, we now briefly review moduli stabilization in the particularly well studied context of IIB string theory compactified on a Calabi-Yau orientifold 
\cite{Kachru:2003aw,Kachru:2003sx,Douglas:2006es,Denef:2008wq,Baumann:2014nda}. At low energies this setup corresponds to a ${\mathcal N}=1$ supergravity theory in four dimensions. The moduli include the complex structure moduli $\xi^\alpha$, the axio-dilaton, and the K\"ahler moduli $T^i$. The resulting  F-term  potential is given by
\be
V_F=e^{K_0}\left(|F|^2-{3}|W_0|^2\right)\,,
\ee
where $F_I=D_I W_0$, $D_I$ is the K\"ahler covariant derivative and the indices $I$ run over all moduli. The leading K\"ahler potential for the K\"ahler and complex structure moduli is given by
\be
K_0=-2\log({\cal V})-\log\left(-i\int\Omega\wedge \bar{\Omega}\right)\,,
\ee
where ${\cal V}$ is the compactification volume measured in string units and $\Omega$ is the holomorphic three-form. For the case of a single K\"ahler modulus the compactification volume is given by ${\cal V}=(T+\bar{T})^{3/2}$. The tree-level holomorphic superpotential $W_0$ depends on the choice of fluxes, but is independent of the K\"ahler moduli, so the potential is of the no-scale form and the F-terms of the K\"ahler moduli do not enter the scalar potential. 
For typical compactification manifolds there are vastly more possible flux choices than there are complex structure moduli, such that we expect a vast number of distinct flux vacua. This is known as the flux landscape and might be able to accommodate an exponentially small vacuum energy density \cite{Bousso:2000xa}.  

Both $\alpha^\prime$ and string loop corrections to the K\"ahler potential break the no-scale structure of the potential. For example, the $\alpha^{\prime 3}$ curvature correction arising via a four-loop correction of the worldsheet $\sigma$-model leads to a contribution to the K\"ahler potential that scales with the inverse  of the compactification volume \cite{Gross:1986iv, Becker:2002nn} .  
Demanding weak coupling requires that the compactification volume is large in string units, ${\cal V}\gg 1$. Famously, the superpotential receives no perturbative corrections, but non-perturbative effects on four cycles of the internal manifold give rise to superpotential terms of the form
\be
W_{\text{np}}=\sum_i A_i e^{-2\pi q^{i}_{\,j} T^j}\,,\label{npw}
\ee
where $A_i\sim \M^3$ are one-loop determinants independent of the K\"ahler moduli, and the entries of the matrix $q^i_{\,j}$ are rational. The terms that appear in (\ref{npw}) may depend on the choice of fluxes. Again, the contributions are only small at large volumes, $2\pi q^{i}_{\,j} T^j\gg 1$. In the same way that complex structure moduli gave rise to the flux landscape, the K\"ahler moduli generically give rise to a vast axion landscape for each flux choice \cite{Bachlechner:2015gwa}.

One  way to arrive at well controlled vacua where all moduli are stabilized is the KKLT scenario \cite{Kachru:2003aw,Kachru:2003sx}. The leading non-perturbative contributions to the superpotential are accidentally competitive with the tree level flux superpotential $W_0$. Let us consider a case where the fluxes are fixed at a high energy scale such that at low energies a single K\"ahler modulus $T$ is the only remaining dynamical field. 
The scalar potential in this simple case becomes
\be\label{kkltpotential}
V={\pi q A e^{-2\pi q \rho}\over \rho^2} \left[W_0+A e^{-2\pi q \rho}\bigg(1+{2\pi q\rho\over 3}\bigg)\right]\,,
\ee
where we defined $T=\rho+i{\chi}$, and set the axion ${\chi}$ to its minimum. For typical values of $W_0$ the scalar potential has a minimum at small volume which is beyond the regime of perturbative control. This is precisely the Dine-Seiberg problem at work. However, for accidentally small values of $W_0$ the potential has a minimum at large volume with small, negative vacuum energy
\be\label{barrier}
V_{\rho_*}\sim-{q^2 A^2\over \rho_*} e^{-4\pi q \rho_* }\,.
\ee
We can see from the form of the potential (\ref{kkltpotential}) that  a highly tuned cancelation of the tree-level  and non-perturbative terms in the superpotential was required for a stable vacuum to arise at large volume. If we are interested in describing a realistic cosmology, we need to find a stable vacuum with positive energy density. Ideally, one might hope to achieve supersymmetry breaking in a way that is decoupled from the K\"ahler moduli stabilization. However, any source of positive energy localized in four dimensions interacts at least with the overall volume of the compactification. At large volume, this corresponds to a coupling of the form
\be
\delta V\sim {D\over {\cal V}^\alpha}\,,
\ee
where $D$ and $\alpha$ are both positive. This coupling can induce a decompactification instability when the source becomes large,  so $D/{\cal V}^\alpha$ is required to be extremely small. Under these conditions we find a weakly coupled vacuum with positive energy density and stable moduli\footnote{The validity of de Sitter vacua in string theory has been critically examined in recent years, see \cite{Baumann:2014nda} for a  review of the related subtleties.}. 

Finally, we are in a position to evaluate the impact of domain walls on moduli stabilization in the KKLT scenario \cite{Johnson:2008vn}.  Consider a landscape parametrized by a field $\phi$, that couples to the K\"ahler modulus $\rho$ via its energy density,
\be\label{}
V(\rho,\phi)={1\over \rho^\alpha} V_\phi(\phi)\,.
\ee
With $\alpha=3$ this would correspond to the form of the F-term potential for the complex structure moduli. Assuming that the modulus is at its stabilized value $\rho_*$, the barrier protecting the vacuum from  decompactfication is given roughly by $V_{\rho_*}$ in (\ref{barrier}). Canonically normalizing the fields and using (\ref{criterion}), we find the critical domain wall tension $\sigma$ that would de-stabilize the modulus,
\be
{\sigma\over \M}\gtrsim {q A\over \sqrt{\rho_*}}e^{-2\pi q \rho }\,.
\ee
Thus, unless the domain wall tension $\sigma$ is exponentially small, the K\"ahler modulus will generically be destabilized by a domain wall. In particular, for domain walls separating vacua with different fluxes, the relevant scale for the domain wall tension is given by that of a five-brane wrapping a three cycle and generically leads to moduli destabilization \cite{Johnson:2008kc,Johnson:2008vn,Aguirre:2009tp}. For an axion landscape the situation is more complicated \cite{Bachlechner:2015gwa}. While the scale of the domain wall tension is set by non-perturbative effects, it is not clear that this will help to maintain stabilized K\"ahler moduli, as in this scenario a large number of axions is required in order to obtain a rich vacuum structure. The high dimensionality  induces additional instabilities so a more sophisticated analysis is required.

\section{Thin Wall Vacuum Transitions}\label{vacuumtransitionss}
In the previous section we discussed a generic runaway instability of domain walls interpolating between de Sitter vacua. Naively, this represents an obstacle for vacuum transitions with thin domain walls. However, it is interesting to ask if there exist  stable transitions between  two de Sitter phases in the absence of any direct domain walls. Before considering this specific problem, we now review the general theory of spatially isotropic thin wall vacuum transitions in 3+1 dimensions. These vacuum transitions are not necessarily the ones relevant in the context of string theory. Domain walls between stabilized vacua and the runaway phase connect phases of different dimensionality. In the runaway phase the four dimensional Newton constant approaches zero, and the validity of the effective field theory terminates. Despite this observation, we will consider the four dimensional effective theory as a proxy for the more complex problem of dynamical compactification. It would be very interesting to study vacuum decay in the higher dimensional   theory, but this is beyond the scope of the present work.

Thin wall vacuum transitions in gravitational theories have been the subject of intense investigation and  most results presented in this section have previously appeared in the literature. However, some of the results are not immediately intuitive, so in this section we present a self-contained discussion of the leading semiclassical dynamics of thin domain walls in the presence of gravity. Although most of our discussion readily applies to vacua with negative energy density, we avoid an explicit review of the related subtleties and focus mostly on de Sitter vacua.

\subsection{The Hamiltonian description of domain walls}
Let us review the Hamiltonian formulation of gravity with spherical symmetry, following \cite{Fischler:1989se,Fischler:1990pk} (see also \cite{Berger:1972pg,Unruh:1976db,Regge:1974zd,Farhi:1986ty,Farhi:1989yr,Kraus:1994by,Ansoldi:1997hz,Ansoldi:2014hta} for related works). Consider $N-1$ spherical domain walls separating $N$  patches of static Schwarzschild-(anti) de Sitter spaces labeled by the index $\alpha=1\dots N$. The metric within each of the patches, defined by a radial coordinate $\hat{r}_{\alpha-1}<r<\hat{r}_\alpha$, is given in static coordinates $T$, $R$ and $\Omega$ by\footnote{We implicitly define $\hat{r}_0\equiv 0$ and $\hat{r}_N\rightarrow \infty$, such that $\hat{r}_\alpha$ is the coordinate of the outer boundary of region $\alpha$.} 
\be\label{staticmetric}
ds_\alpha^2=-A_\alpha(R) dT^2+A_\alpha^{-1}(R) dR^2+R^2 d\Omega_2^2\,,
\ee
where $d\Omega_2^2$ is the metric on a unit two-sphere.  While the spacetime is static within each region, the domain walls are dynamic, so let us consider the most general spatially isotropic metric in 3+1 dimensions,
\be\label{sphericalmetric}
ds^2
=-N^{t2}(t,r)dt^2+L^2(t,r)[dr+N^r(t,r) dt]^2+R^2(t,r)d\Omega_2^2\,.
\ee
The lapse $N^t(t,r)$ and shift $N^r(t,r)$ are non-dynamical and set the gauge, while $R(t,r)$ corresponds to the radius of curvature of the two-sphere at coordinates $t$ and $r$. From now on we will mostly omit the explicit dependence on the coordinates to simplify the notation. Consider a scalar field theory  coupled to gravity on a spacetime region $\mathcal M$ with time-like boundary $\partial {\mathcal M}$. The action is given by\footnote{In this work $\M^{-2}=8\pi G$ is the reduced Planck mass.}
\be\label{ehcation}
S={\M^2\over 2} \int_{\mathcal M} d^4x~\sqrt{g} {\cal R}+\M^2\int_{\partial {\mathcal M}}d^3y~ \sqrt{h}{\cal K}+ \int_{\mathcal M} d^4x~\sqrt{g} {\mathcal L}_m +S_\text{b.t.}\,,
\ee
where $h_{ij}$ and ${\mathcal K}$ are the induced metric and extrinsic curvature on $\partial {\mathcal M}$, respectively, and $g_{\mu\nu}$ is the four dimensional metric on $\mathcal M$. The last contribution $S_\text{b.t.}$ denotes boundary terms that may be necessary in order for Hamilton's equations to hold. For simplicity, we will assume that any intrinsic dynamics of the matter Lagrangian occur over time scales much shorter than any  scale relevant for the dynamics of the spherically symmetric domain walls, so we can approximate the matter Lagrangian ${\mathcal L}_m\approx-\rho(r)$ away from domain walls. We further assume that different phases are separated by a thin wall with surface energy density $\sigma$ and  tension $\xi$. The domain wall separates two (approximate) vacua of a scalar field theory, such that the surface tension equals the energy density of the domain wall. The energy momentum tensor is then given by \cite{Blau:1986cw}
\be
T^{\mu\nu}=-\sum_{\alpha=1}^{N-1}\sigma_\alpha h_\alpha^{\mu\nu}\delta(r-\hat{r}_\alpha)-\sum_{\alpha=1}^{N}g^{\mu\nu}\rho_{\alpha}\,,
\ee
where $\sigma_\alpha=\xi_\alpha$ and $\hat{r}_\alpha$ are the is the surface tension and radial coordinate of the domain wall separating regions $\alpha$ and $\alpha+1$. We denote the metric at the shell by $h_{\alpha}^{\mu\nu}$. The energy density $\rho_\alpha$ is only non-vanishing within region $\alpha$. We denote all quantities evaluated at a domain wall with a hat  and an index specifying the domain wall.
To simplify the analysis, we bring the action into first-order form. Deferring the derivation to Appendix \ref{app1}, the dynamical terms are given in terms of canonical variables as\footnote{We denote partial derivatives with respect to $r$ and $t$ with primes and dots, respectively.}
\cite{Fischler:1990pk}
\bea\label{action2}
S&=& \int dtdr~\left( \pi_L\dot{L}+\pi_R \dot{R}-N^t {\cal H}_t-N^r {\cal H}_r\right)+\sum_{\alpha=1}^{N-1} S_{\text{Shell},\,\alpha}\,,\\
S_{\text{Shell},\,\alpha}&=&\int dt \,\hat{p}_\alpha\dot{\hat{r}}_\alpha+\int d\hat{R}_\alpha\,{\hat{R}_\alpha\eta_{\pi}\over G}\log\left( {{\hat{R}^\prime}_{\alpha,+}-\sqrt{{\hat{R}^{\prime2}}_{\alpha,+}-\hat{L}^2_\alpha\hat{A}_{\alpha+1}}\over {\hat{R}^\prime}_{\alpha,-}-\sqrt{{\hat{R}^{\prime2}}_{\alpha,-}-\hat{L}^2_\alpha\hat{A}_{\alpha}} }\sqrt{\hat{A}_\alpha\over \hat{A}_{\alpha+1}}\right),\nonumber
\eea
where $\eta_{\pi}=\text{sgn}(\pi_{L})$, and the second term in  $S_{\text{Shell},\,\alpha}$ arises due to a possible discontinuity of $R^\prime$ at the shells \cite{Fischler:1989se,Fischler:1990pk,Kraus:1994by}. The Hamiltonian densities are given by
\bea\label{hamilt}
&&{\cal H}_t={G L \pi_L^2\over 2 R^2}-{G\over R} \pi_L\pi_R+{\left({2 R R^\prime\over L}\right)^\prime -{R^{\prime2}\over L}-L\over2 G}+4\pi L R^2 \rho(r)+\sum_{\alpha=1}^{N-1}\delta(r-\hat{r}_\alpha) \sqrt{{\hat{p}_\alpha^2\over L^2}+m_\alpha^2}\nonumber,\\
&&{\cal H}_r=R^\prime \pi_R-L\pi_L^\prime-\sum_{\alpha=1}^{N-1}\delta (r-\hat{r}_\alpha) \hat{p}_\alpha\,,
\eea
and $\pi_L$, $\pi_R$ and $\hat{p}$ are  momenta conjugate to the canonical variables $L$, $R$ and $\hat{r}$, while $N^r$ and $N^t$  appear as non-dynamical Lagrange multipliers that impose the secondary Hamiltonian constraints. The conjugate momenta are related to the velocities by
\be
\pi_L={N^r R^\prime-\dot{R}\over G N^t}R\,,~~
\pi_R={(N^r L R)^\prime - \partial_t (L R)\over GN^t } \,,~~
\hat{p}_\alpha={m_\alpha \hat{L}^2_\alpha (\hat{N}^r_\alpha+\dot{\hat{r}}_\alpha)\over \sqrt{\hat{N}^{t2}_\alpha-\hat{L}_\alpha^2(\dot{\hat{r}}_\alpha+\hat{N}_\alpha^r)^2}}\,,
\ee
where $m_\alpha=4\pi \sigma_\alpha \hat{R}^2_\alpha$. The action does not contain kinetic terms for the shift and lapse, which implies the secondary constraints
\be\label{const0}
{\cal H}_{t}={\cal H}_{r}=0\,.
\ee
By considering these constraints at the location of each of the shells, we  obtain the required $2(N-1)$ equations that determine the classical domain wall dynamics, in addition to the $2N$ constraints that determine the spatial geometry between domain walls.
For constant energy densities within the domains we can rewrite the constraint equations as
\be
\pi_{L,\alpha}=\eta_{\pi} {R\over G}\sqrt{{R^{\prime2}\over L^2} -A_{\alpha}}\,,~~~\pi_{R,\alpha}={L\over R^\prime} \pi_{L,\alpha}^\prime\,.\label{constraints}
\ee
After imposing the constraints (\ref{constraints}) on the system, we find an explicit expression for the full action,
\be
S=S_{\text{Space}}+\sum_{\alpha=1}^{N-1} S_{\text{Shell},\,\alpha}\,,
\ee
where
\bea\label{spacepart12}
S_{\text{Space}}=\sum_{\alpha=1}^N\int_{\hat{r}_{\alpha-1}+\epsilon}^{\hat{r}_{\alpha}-\epsilon} dr~{R \eta_{\pi}\over G}\left(\sqrt{{R^{\prime2}}-L^2 A_\alpha} - {R^\prime } \text{arccosh}\left[{{R^\prime}  \over\sqrt{L^2A_\alpha}}\right]\right)\,.
\eea

\begin{figure}
  \centering
  \includegraphics[width=1\textwidth]{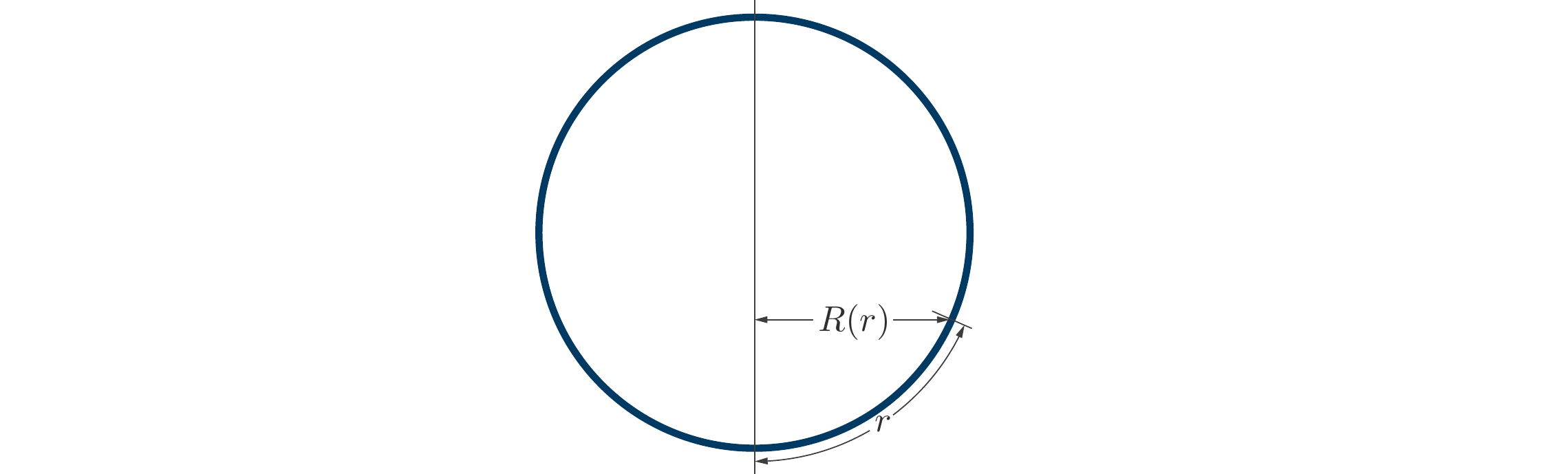} 
  \caption{\small Illustration of a spatial slice of de Sitter space. The quantity $R$ is the physical radius of curvature, while the coordinate $r$ parametrizes the spatial slice. \label{desitterfigure}}
  \end{figure}

Let us pause for a moment and consider a single domain of positive energy density and vanishing mass parameter, so the metric is given by (\ref{staticmetric}) with $A=1-H^2 R^2$. We pick coordinates where $L=1$, $N^r=0$ and $N^t=1$. The constraint equation then has the  solution
\be
R={H}^{-1}\sin({H} r)\,.
\ee
As expected, this simply corresponds to a spatial slice of de Sitter space. The corresponding three-geometry is illustrated in Figure \ref{desitterfigure}.

We now consider the junction conditions at the domain wall locations by integrating the constraint equations across the defects and noting that both $R$ and $L$ are continuous across the walls. We denote the derivatives just inside and outside the domain wall $\alpha$ by $\hat{R}^\prime_{\alpha,\pm}= R^\prime|_{r=\hat{r}_\alpha\pm\epsilon}$. Using this notation we can integrate (\ref{const0}) in the vicinity of the shell located at $\hat{r}_\alpha$ to find 
\bea\label{junctionfulltext}
\hat{\pi}_{L,\alpha,+}-\hat{\pi}_{L,\alpha,-}&=&-{\hat{p}_\alpha\over \hat{L}_\alpha}\,,\\\label{junction22}
\hat{R}^\prime_{\alpha,+}-\hat{R}^\prime_{\alpha,-}&=&
-{G\over \hat{R}_\alpha }\sqrt{{\hat{p}_\alpha^2}+m_\alpha^2\hat{L}_\alpha^2 }
\,.
\eea
In the rest frame of a domain wall the momentum of the wall vanishes, $\hat{p}_\alpha=0$, and we find the simple junction conditions 
\bea\label{rpsimplegauge}
\hat{R^\prime}_{\alpha,\pm}&=&{A_\alpha(\hat{r}_\alpha)-A_{\alpha+1}(\hat{r}_\alpha) \over 2\kappa \hat{R}_\alpha}\hat{L}_{\alpha}\mp{1\over 2} {\kappa  \hat{L}_{\alpha}\hat{R}_\alpha}\,,\\
\hat{\pi}^2_{L,\alpha,-}=\hat{\pi}_{L,\alpha,+}^2&=&{\left(A_\alpha(\hat{r}_\alpha)-A_{\alpha+1}(\hat{r}_\alpha)\right)^2\over  4 G^2 \kappa_\alpha^2}-{A_\alpha(\hat{r}_\alpha)+A_{\alpha+1}(\hat{r}_\alpha)\over 2 G^2}\hat{R}_\alpha^2+{\kappa_\alpha^2 \hat{R}_\alpha^4\over 4 G^2}\,,
\eea
where we introduced the rescaled domain wall tension $\kappa_\alpha=4\pi \sigma_\alpha G$. These junction conditions determine the dynamics of a given domain wall \cite{Israel:1966rt}. Evaluating an angular component of the wall's extrinsic curvature shows that the sign of $\hat{R}^\prime$ determines whether the domain wall is curved towards what we called the interior, or the exterior \cite{Blau:1986cw},
\be\label{curvatureqe}
\hat{K}_{\theta\theta,\,\pm}\propto\hat{R}^\prime_\pm\,,
\ee
so that sign will be important to develop a good intuition for the domain wall dynamics. Remember that the definition of ``interior'' and ``exterior'' is quite arbitrary and we should not be surprised to find both possible curvatures.

In the rest frame the coordinate $t$ corresponds to the proper time of an observer and we now pick the coordinate $r$ such that $L=1$. The constraint equation for $\pi_L$ gives a simple relationship between the velocity $\dot{R}$ and the more abstract quantity $R^\prime$ that determines the extrinsic curvature,
\be\label{rprelationship}
{\hat{R}^{\prime2}_{\alpha,+}}=\dot{\hat{R}}_\alpha^2+A_{\alpha+1}(\hat{r}_\alpha)\,,~~{\hat{R}^{\prime2}_{\alpha,-}}=\dot{\hat{R}}_\alpha^2+A_{\alpha}(\hat{r}_\alpha)\,.
\ee
With (\ref{rpsimplegauge}) we see that $\dot{\hat{R}}$ is continuous across the shell: it corresponds to the velocity of the wall and should agree for observers traveling on either side, but close to the wall. We can use (\ref{rprelationship}) and  the normalization of the four-velocity to find the static coordinate time $T$ of (\ref{staticmetric}) in terms of the proper time $t$ at the wall,
\be\label{rpeq}
\dot{\hat{T}}_{\alpha,\,-}=\pm{\hat{R}^{\prime}_{\alpha,\,-}\over A_{\alpha}(\hat{r}_\alpha)}\,,~~\dot{\hat{T}}_{\alpha,\,+}=\pm{\hat{R}^{\prime}_{\alpha,\,+}\over A_{\alpha+1}(\hat{r}_\alpha)}\,,
\ee
where the sign of $\dot{\hat{T}}$ is set by convention. In the next subsection we will use the relations (\ref{rprelationship}) and (\ref{rpeq}) to obtain the domain wall dynamics both in terms of static coordinate time and  proper time along a trajectory.

\subsection{Classical domain wall dynamics}\label{classicaldynamcis}
We now proceed to discuss the classical dynamics of a single domain wall in its rest frame. The generalization to multiple domain walls is straightforward, but to obtain simple constraint equations a different gauge is needed for each wall. 
We consider spherical regions of Schwarzschild-(anti) de Sitter spacetimes with energy density $\rho_{\alpha}=3H^2_{\alpha}\M^2$ in region $\alpha$. The metric is given in  (\ref{staticmetric}) with
\be
A_{\alpha}(R)=1-H_{\alpha}^2 R^2-{2 G M_{\alpha}\over R}\,.
\ee
We can rewrite the constraint equations (\ref{rpsimplegauge}) and (\ref{rprelationship}) to find 
the asymptotic mass parameter in the exterior region, 
\be\label{masseq}
M_{\alpha+1}={{H}_{\alpha}^2-{H}_{\alpha+1}^2\over 2G}\hat{R}_{\alpha}^3-{\kappa_{\alpha}^2 \over 2 G} \hat{R}_{\alpha}^3+m_\alpha\,\text{sgn}(\hat{R^\prime}_{\alpha,\,-})\sqrt{1-{H}^2_{\alpha}\hat{R}_{\alpha}^2-{2 G M_{\alpha}\over \hat{R}_{\alpha}}+\dot{\hat{R}}_{\alpha}^2}+M_\alpha\,,
\ee
where $m_\alpha=4\pi \sigma \hat{R}_\alpha^2$. Each of these terms has a simple and intuitive interpretation. The first term is the contribution due to the vacuum energy density, the second term constitutes the gravitational surface-surface interaction term, and the third term is due to the energy of the shell. As one would expect, the latter term always drives the domain wall towards a smaller radius of curvature. Since the radius of curvature can either increase or decrease with the radial coordinate $r$, the sign of this term depends on $R^\prime$. 
\begin{figure}
  \centering
  \includegraphics[width=1\textwidth]{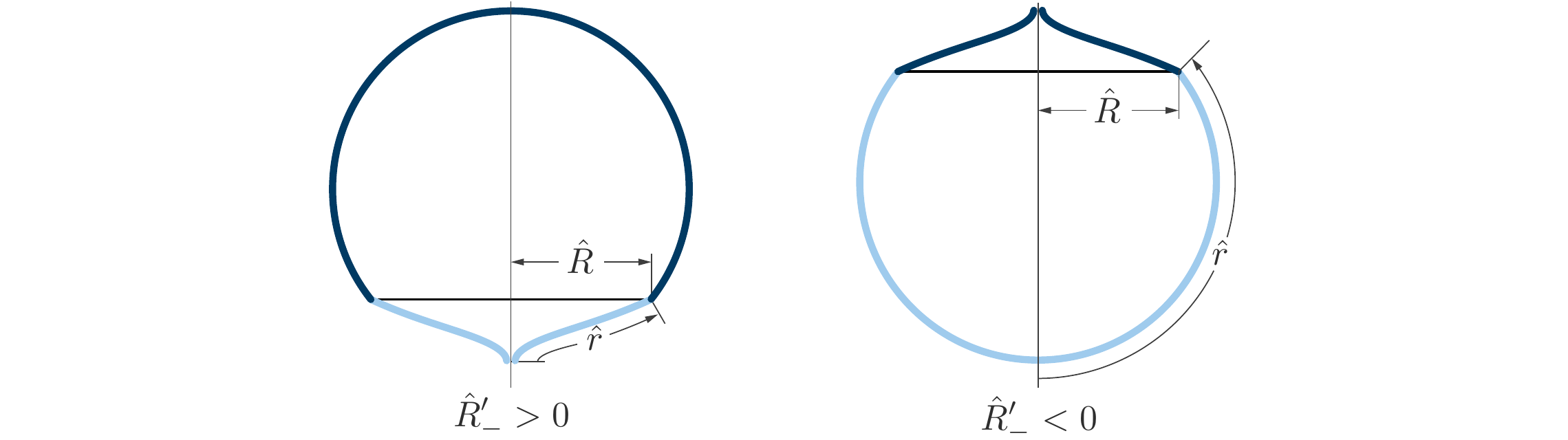}  
    \caption{\small Spatial slices of two de Sitter phases joined by a domain wall. The interior and exterior geometries are shown in light and dark lines, respectively. The left figure shows a bubble with a true vacuum interior, while the right shows a bubble with a false vacuum interior. The geometries are physically identical and related by a coordinate change.   \label{spatialbubbles}}
  \end{figure}
To illustrate this feature, we can integrate (\ref{rpsimplegauge}) and recover the spatial geometry at a classical turning point, where $\pi_L=0$. The resulting spatial geometries are shown in Figure \ref{spatialbubbles} for physically identical configurations but different signs of $R^\prime$. The domain wall tension always seeks to decrease the radius of curvature $R$. When the physical radius increases with the coordinate $r$, the tension forces the domain wall to smaller values of $r$. In this case the contribution from the domain wall tension to the asymptotic outside mass is positive. On the other hand, when the radius of the three sphere decreases with $r$ at the location of the wall, the situation is reversed and the domain wall tension contributes with a negative sign to the outside asymptotic mass. The quantity $R^\prime$ at the domain wall is determined by (\ref{rpsimplegauge}) and can be written as
\bea\label{rpsds}
\hat{R}^\prime_{\alpha,\,\pm}= \frac{2 G (M_{\alpha+1}-M_{\alpha})+\hat{R}_\alpha^3 \left({H}_{\alpha}^2-{H}_{\alpha+1}^2\mp \kappa_\alpha ^2\right)}{2 \kappa_\alpha  \hat{R}_\alpha^2}\,.
\eea
The constraint equations determine the classical dynamics of  domain walls separating static spacetimes, so in the absence of collisions there are no dynamic interactions and it is sufficient to consider the walls independently. In the following we therefore constrain the discussion to a single domain wall separating two Schwarzschild-(anti) de Sitter spacetimes, labeled by indices $\alpha={\cal B}\,,{\cal A}$. This problem has been intensely studied in the literature, so we only present some of the main results in this section and refer to the references for details \cite{Blau:1986cw,Aguirre:2005nt}.
\begin{figure}
\centering
\includegraphics[width=1\textwidth]{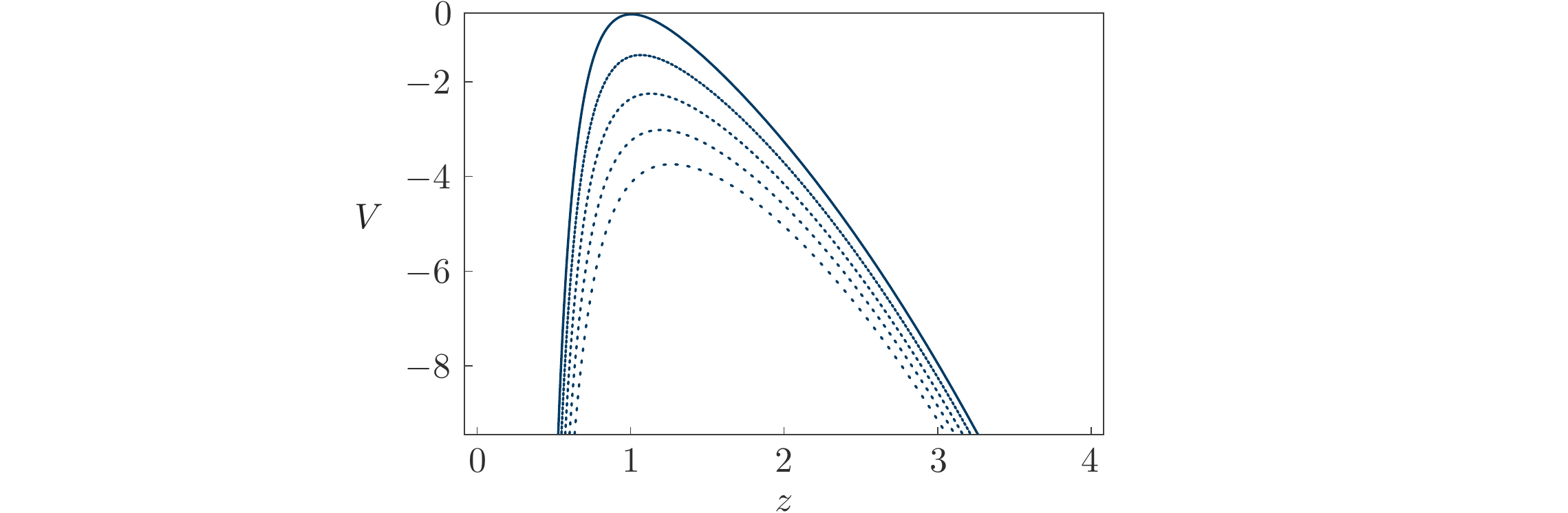}
\caption{\small Effective potential $V(z)$ as a function of the rescaled radial coordinate $z$ for varying parameters $-2\le\gamma\le2$.}\label{potentialexample}
\end{figure}

To set the notation, we discuss the dynamics of a spherically symmetric domain wall at a radial coordinate $\hat{r}$ separating two  static metrics of the form (\ref{staticmetric}) where
\bea\label{metricfactors}
A_{\cal B}=1-{H}^2_{\cal B} R^2-{2 G M_{\cal B}\over R}\,,~~~\text{for $r<\hat{r}$}\,,\nonumber\\
A_{\cal A}=1-{H}^2_{\cal A} R^2-{2 G M_{\cal A}\over R}\,,~~~\text{for $r>\hat{r}$}\,,
\eea
and ${\cal B}$ and ${\cal A}$ denote the inside and outside regions, respectively.  We can solve the constraint equation (\ref{masseq}) to find an expression that is quadratic in $\dot{\hat{R}}^2$ and has the form of an energy conservation equation. To simplify the expression we define a new radial coordinate $z$ by
\be
z^3\equiv \frac{H_+^2}{2 G |M_{\cal B}-M_{\cal A}|} \hat{R}^3\,,~~H_+^4=\left(\kappa ^2+{H}_{\cal B}^2+{H}_{\cal A}^2\right)^2-4 {H}_{\cal B}^2 {H}_{\cal A}^2\,.
\ee
This leaves the constraint equation in a particularly simple form of an energy conservation equation for a particle moving under the influence of an effective  potential $V(z)$,
\be
{4\kappa^2\over H_+^4}\left({\partial z\over \partial t}\right)^2+V(z)=E\,,~~V(z)=-\left(z^2 +{\gamma\over z}+{1\over z^4}\right)\,,
\ee
where the real, negative energy and the parameter $\gamma$ are given by
\be
|E|^{3/2}=\frac{8 \kappa ^3}{  2G| M_{\cal A}-M_{\cal B}|H_+^4}\,,~~\gamma={2\,\text{sgn}(M_{\cal B}-M_{\cal A})\over{H}_+^2}\left({H}_{\cal B}^2-{H}_{\cal A}^2+{M_{\cal B}+M_{\cal A}\over M_{\cal B}-M_{\cal A}}\kappa^2\right)\,.
\ee
We illustrate the effective potential in Figure \ref{potentialexample}. For fixed asymptotic masses the parameter $\gamma$ determines whether the gravitational interaction of the domain wall tension or the energy density are dominant in the dynamics. If the mass parameter vanishes on either side of wall, the small tension or weak gravity limit corresponds to $\gamma\approx -2$.  The dynamics within the potential are determined by the constant of motion $E$. The effective potential has a maximum at a finite coordinate $z_{\text{max}}$, where $V_{\text{max}}=V(z_\text{max})\le 0$, such that there exist both bound and unbound solution when the masses are finite. When both asymptotic masses vanish the only ``bound'' solutions are static configurations at $\hat{R}=0$.

\subsection{Dynamics of false vacuum bubbles}\label{dSSdynamics}
\begin{figure}
  \centering
  \includegraphics[width=1\textwidth]{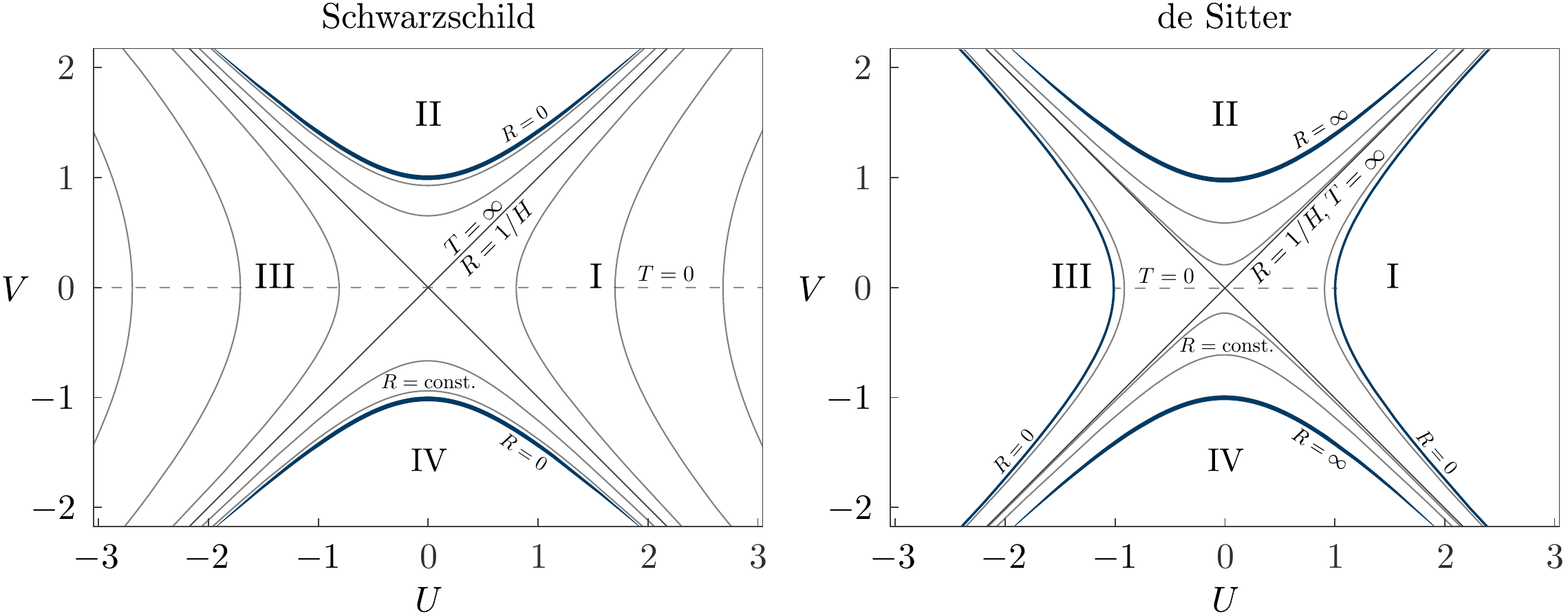}
  \caption{\small Spacetime diagrams of Kruskal-Szekeres coordinates for Schwarzschild (left) and Gibbons-Hawking coordinates of de Sitter space (right). The solid gray lines illustrate constant radius surfaces, while the dashed line illustrates the $T=0$ surface. Static coordinate time $T$ is increasing upwards/downwards in regions I/III.}\label{desitterschwarzschild}
\end{figure}
We now turn to the discussion of domain walls separating a spherical patch of de Sitter space from a Schwarzschild spacetime. While the generalization to arbitrary masses and energy density densities is conceptually straightforward, this particular special case will be of most interest to our remaining discussion and is simple to illustrate.  

In order to illustrate the causal structure of the trajectories we can change to new coordinates $U$ and $V$ that are smooth everywhere, and implicitly denote Gibbons-Hawking coordinates for the de Sitter region and Kruskal-Szekeres coordinates for the Schwarzschild region. The coordinates are explicitly defined in Appendix \ref{app2}. We illustrate both spacetimes in Figure \ref{desitterschwarzschild}. In these coordinates light  travels along $45$ degree angles, and the spacetimes are divided into four regions  by the Schwarzschild and de Sitter horizons at  $|U|=|V|$. The metrics in the new coordinates become
\bea
ds^2_{\text{KS}}&=&{32 (G M)^3\over R}e^{-R/2GM} (-dV+dU^2)+R^2 d\Omega^2_2\,,\\
ds^2_{\text{GH}}&=&\left({1+{H} R\over {H}}\right)^2(-dV^2+dU^2)+R^2 d\Omega^2_2\,.
\eea
\begin{figure}
  \centering
  \includegraphics[width=1\textwidth]{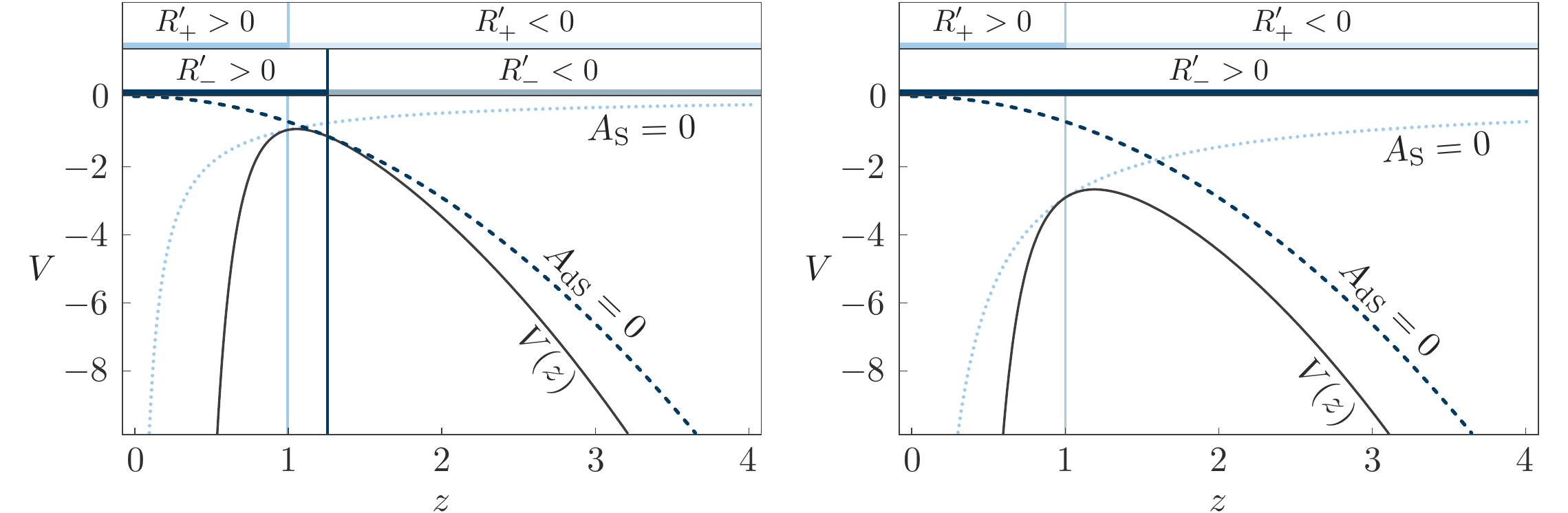}
  \caption{\small Effective potential $V(z)$ as a function of rescaled radial coordinate $z$, for $\kappa<H$ (left) and $\kappa>H$ (right) for a de Sitter/Schwarzschild domain wall. The horizons are shown by dashed and dotted lines in the diagram and the sign of the extrinsic curvature on both sides of the domain walls is illustrated on the top.}\label{figurefull2}
\end{figure}

One convenient feature of Kruskal-Szekeres and Gibbons-Hawking coordinates is that with (\ref{rpeq}) the change in polar angle along a given trajectory is directly proportional to the change of the radius of curvature with $r$,
\be
\partial_t\, \text{arctan}\left({V(t)\over U(t)}\right)\propto A\, \partial_t T\propto \pm R^\prime(t)\,.
\ee
We can chose a convenient convention for how $R^\prime$ is related to the change in coordinate time with proper time in (\ref{rpeq}) by picking opposite signs for the de Sitter and Schwarzschild regions,
\be\label{tderivative}
\dot{T}_{{\text{dS}}}=-{R^{\prime}\over A_{{\text{dS}}}(R)}\,,~~~\dot{T}_{{\text{S}}}={R^{\prime}\over A_{{\text{S}}}(R)}\,.
\ee
This coordinate definition implies that a light ray crossing a given time-like trajectory appears as propagating in the same direction in either diagram. The opposite sign choice stems from the fact that for the Schwarzschild diagram an increase in the coordinate $|U|$ ($|V|$) corresponds to an increase (decrease) in the radial coordinate R in region I/III (II/IV), while the reverse is true for the de Sitter diagram. While this choice has no physical consequences it allows for a simpler interpretation of the coordinate diagrams and we can immediately determine the sign of the extrinsic curvature of the domain wall from the coordinate diagram. For example, a trajectory where the domain wall curvature is positive on both sides would appear in region I of the Schwarzschild diagram, with an increasing polar angle, while it would appear in region III of the de Sitter diagram with a decreasing polar angle\footnote{As we shall see, this kind of trajectory could correspond to an expanding true vacuum bubble.}. In both diagrams the trajectory moves upwards with increasing proper time. 
\begin{table*}
\begin{center}
\begin{tabular}{@{}p{2.5cm} @{}p{3.5cm}@{}p{3.5cm}  @{}p{4.8cm} }\\\toprule
Type& de Sitter & Schwarzschild &Conditions\\\midrule
Bound&III&IV - I - II& $E<V_{\hat{R}^\prime_+=0}<V_\text{max}$ \\
Bound&III&IV - III - II& $V_{\hat{R}^\prime_+=0}<E<V_\text{max}$ \\
Unbound&III-II&IV - III& $V_\text{max}<E$ \phantom{$V_{\hat{R}^\prime_+}$}\\
Unbound& IV - III - II&III& $V_{\hat{R}^\prime_-=0}<E<V_\text{max}$,\,~$\kappa<H$ \\
Unbound&IV - I - II&III& $E<V_{\hat{R}^\prime_-=0}<V_\text{max}$ \\
\bottomrule\\
\end{tabular}
\caption{\label{dSStrajectories}Summary of all possible trajectories of a de Sitter/Schwarzschild domain wall. Depending on the constant of motion $E$ and the initial conditions the domain wall trajectories are bound, or unbound, and pass through different regions of the conformal diagrams.}
\end{center}
\end{table*}

For the case of a de Sitter (inside)/Schwarzschild (outside) domain wall the new, rescaled radial coordinate $z$ simplifies and we have
\be\label{zdss}
z^3= \frac{{H}_+^2}{2 G M_{\cal A}} \hat{R}^3\,,~~{H}_+^2={4{H}_{{\cal B}}\over 2-\gamma}\,,~~\gamma={2}{\kappa^2-{H}_{{\cal B}}^2\over \kappa ^2+{H}_{\cal B}^2}\,,~~ E=-\frac{4 \kappa ^2}{  (2G M_{\cal A}) ^{2/3}{H}_+^{8/3}}\,.
\ee
Here we see that $\gamma\approx -2$ when the rescaled domain wall tension $\kappa=4\pi G \sigma$ is small compared to the Hubble scale.  There are two main quantities that we are interested in when considering the domain wall dynamics: the change of the radius of curvature with radial coordinate $\hat{R}^\prime$, and the location of horizons. 
We can express these quantities in terms of the new radial coordinate $z$ as
\be\label{rprimeandstuff}
{\hat{R}^\prime_{-}}={2+\gamma z^3\over 2\sqrt{|E|} z^2}\,,~~{\hat{R}^\prime_{+}}={1-z^3\over \sqrt{|E|} z^2}\,, ~~z_{A_{{\cal B}}=0}={\sqrt{|E|\over 1-\gamma^2/4}}\,,~~z_{A_{{\cal A}}=0}={2+\gamma\over |E|}\,.
\ee
We immediately see with (\ref{curvatureqe}) that the wall's extrinsic curvature is positive for small radii. When the domain wall tension is dominant ($\gamma>0$), the extrinsic curvature on the de Sitter interior is always positive. To illustrate the full dynamics we  show the potential in Figure \ref{figurefull2}.
From this figure we can determine all possible domain wall trajectories. There are bound solutions and unbound solutions. The bound solutions emerge from vanishing size, and collapse after bouncing off the potential. The unbound solutions recede from infinite size, approach a finite radius and expand again. In \cite{Farhi:1986ty} it was shown that the unbound solutions are not buildable by classical dynamics: they always contain a singularity in their past. However, it is possible that a non-singular, bound solution tunnels through the potential barrier and emerges as an unbound solution. All possible classical domain wall trajectories are summarized in Table \ref{dSStrajectories}. 

\begin{figure}
\centering
\includegraphics[width=1\textwidth]{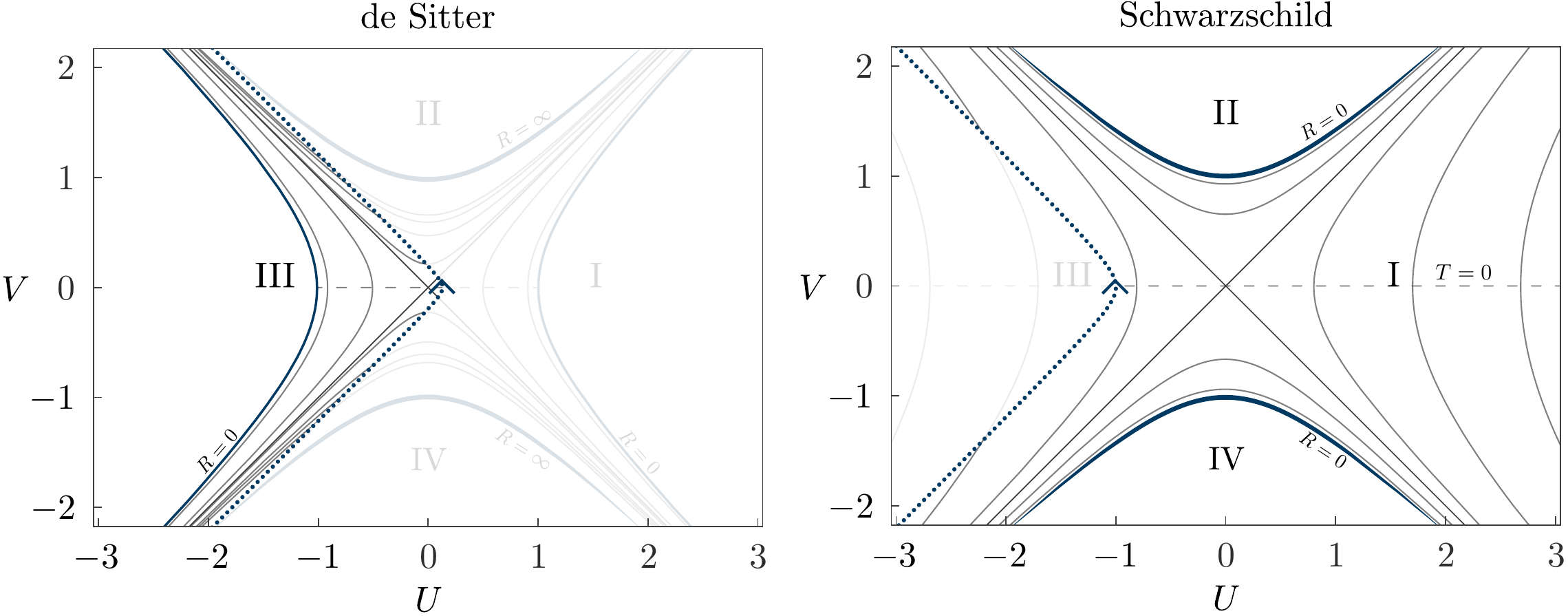}
\caption{\small\label{dsSsols}Illustration of the unbound trajectory of a de Sitter/Schwarzschild domain wall that traverses the regions IV-I-II in the de Sitter diagram and is contained in region III of the Schwarzschild diagram. The spacetime diagram is not applicable in the faded region.}
\end{figure}
Note that low mass, unbound solutions have negative extrinsic curvature when they come to rest and start expanding, i.e. the radius of curvature decreases as we pass from the de Sitter to the Schwarzschild region. These solutions pass through region III of the Schwarzschild diagram and region I of the de Sitter diagram. 
This means that the bubble contains more than half of a spatial slice of de Sitter space when it comes to rest, so the de Sitter region is causally protected from extinction. To an observer in the Schwarzschild phase, who experiences a reversed definition of inside and outside,  the situation would appear reversed: both extrinsic curvatures are  positive for them, i.e. the radius of curvature increases as they approach the domain wall from their inside, which corresponds to the well known CDL domain wall evolution.
An example of an unbound trajectory, as seen both in the Schwarzschild and de Sitter coordinates, is shown in Figure \ref{dsSsols}.

The  dynamics of a bubble containing a de Sitter phase inside a Schwarzschild region are identical to the dynamics of a bubble of Schwarzschild phase inside a de Sitter region, with the exception that the coordinate $r$, which determines the direction from inside to outside, is redefined. This reflects the arbitrariness of our definition of what we called inside ($r<\hat{r}$) and outside ($r>\hat{r}$), and is immediately clear when recalling the spatial geometry of the two cases in Figure \ref{spatialbubbles}. In particular, as we reverse the definition of inside and outside all derivatives with respect to $r$ change sign, such that with (\ref{tderivative}) the trajectories will appear on opposite signs of the Kruskal-Szekeres coordinate diagram. Therefore, the present discussion of a false vacuum bubble already captures the full dynamics of true vacuum bubbles. In Appendix \ref{app2} we provide a brief explicit discussion of the Schwarzschild (inside) - de Sitter (outside) domain wall dynamics for reference.

\subsection{Semiclassical transition probabilities}\label{vactransitions}
Studying dynamics revealed both bound and unbound trajectories that are separated by a classical potential barrier. A classical solution cannot penetrate the effective potential, but gets reflected at the turning point of the potential where $\pi_L=0$. A natural question to ask is whether transmission through the barrier is possible in a quantum theory. This question was first asked by Coleman and de Luccia for true vacuum bubbles \cite{Coleman:1980aw}, and by Farhi and Guth for disconnected false vacuum bubbles \cite{Farhi:1986ty,Farhi:1989yr}. Recall that our definition of the bubble interior is arbitrary for de Sitter vacua, so the classical dynamics are equivalent for true and false vacuum bubbles. This invariance is manifest in the Hamiltonian formulation which allows to approach the nucleation of true and false vacuum bubbles within a unified framework. 
We proceed to review the leading semiclassical evolution of domain walls in the Hamiltonian framework to find explicit tunneling trajectories and transition probabilities \cite{Fischler:1989se,Fischler:1990pk,Kraus:1994by}.

In order to employ Dirac quantization, we impose the constraints on the wave function $\psi$, which yields the Wheeler-DeWitt equation \cite{DeWitt:1967yk}
\be
\pi_{N^t,N^r}\psi={\cal H}_{r,t}\psi=0\,.
\ee
The primary constraints demand that the wave function is independent of the gauge choices $N^t$ and $N^r$, and depends only  on the spatial geometry. In order to avoid the ambiguities of quantizing a gravitational theory we impose the classical equations of motion and expand the wave function in the WKB approximation as
\be\label{wkbwavefunction}
\psi=e^{-{i} { S/\hbar}+\mathcal{O}(\hbar^0)}\,,
\ee
such that to leading order in $\hbar$ the exponential dependence of the wave function is related to the classical action $S$. Remember that this prescription only holds for the leading semiclassical approximation. In this work we restrict ourselves to the leading contribution, and perturbations around the leading contribution to the path integral are prohibited. We will mostly be interested in the transition rate from some initial state\footnote{Remember that in the framework of canonical quantization the states correspond to spatial geometries.} ${\mathcal I}$ to a final state ${\mathcal F}$. The transition probability, derived from the transmission coefficient, is given by 
\be\label{transitionprob}
{\cal P}_{{\mathcal I}\rightarrow {\mathcal F}}= e^{-\eta_{\pi}B}= \exp\left(-{2i \int_{{\mathcal I}}^{{{\mathcal F}}} d S}\right)\,.
\ee
The sign $\eta_\pi$ in the exponent of what we will interpret as a probability is potentially ambiguous. We proceed to define a tunneling exponent $B$ with fixed sign and leave the yet undetermined sign in the definition of the action $S$, see (\ref{action2}).

We should emphasize at this point that the initial state is  not a pure de Sitter phase, but corresponds to a classical turning point of a bound or unbound domain wall trajectory.
Only in the  massless limit, where the turning point of the bound domain wall trajectory corresponds to $\hat{R}=0$, is the bound classical turning point  an empty de Sitter space. In order to interpret the tunneling probability ${\cal P}$ as a transition rate $\Gamma$, the bound turning point of the trajectory should arise at some frequency. In the massless limit a de Sitter phase constantly satisfies the required initial conditions, while the massive case may require a thermal fluctuation. Note that even though the radius of curvature of the domain wall vanishes at the bound turning point in the massless configuration, the curvature scalar remains finite everywhere during the bubble nucleation process. 
In the vanishing mass limit we can therefore use the transition probability (\ref{transitionprob}) to estimate the leading exponential dependence of the the vacuum transition rate, $\Gamma\sim {\cal P}$.

At first sight the situation is somewhat precarious: we obtained the transmission coefficient for a massless energy eigenstate, which of course is time-independent. In quantum mechanics, this dilemma is resolved by considering the non-perturbative contribution to the energy due to all possible tunneling trajectories, which yields a small imaginary part for the energy of the metastable state. We defer a thorough derivation of the decay rate to a subsequent publication \cite{BDEMtoappear}, and merely present a heuristic argument in this work. 
In particular, the sign of the action along the relevant tunneling trajectory in (\ref{transitionprob}), $\eta_\pi$, is yet undetermined, and potentially ambiguous. In principle, the sign is determined by the WKB matching conditions, but since we dropped all time-dependence it is not obvious whether a given mode is ingoing or outgoing. The sign is known in the $G\rightarrow 0$ limit, where $\eta_{\pi}=+1$ \cite{Coleman:1980aw}. In the presence of causally disconnected regions this limiting case may not be instructive, and there have been multiple proposals for fixing the sign in various contexts, \cite{Hartle:1983ai,Linde:1983mx,Vilenkin:1984wp,Vilenkin:1986cy}. Naively we might expect that the decay rate is small whenever the action of a classically forbidden trajectory becomes large, which fixes the sign to be $\eta_{\pi}=\text{sgn}(B)$ and coincides with the tunneling wave function proposal by Vilenkin \cite{Vilenkin:1984wp,Vilenkin:1986cy}. A careful derivation of this result is presented in \cite{BDEMtoappear}.

\begin{figure}
  \centering
  \includegraphics[width=1\textwidth]{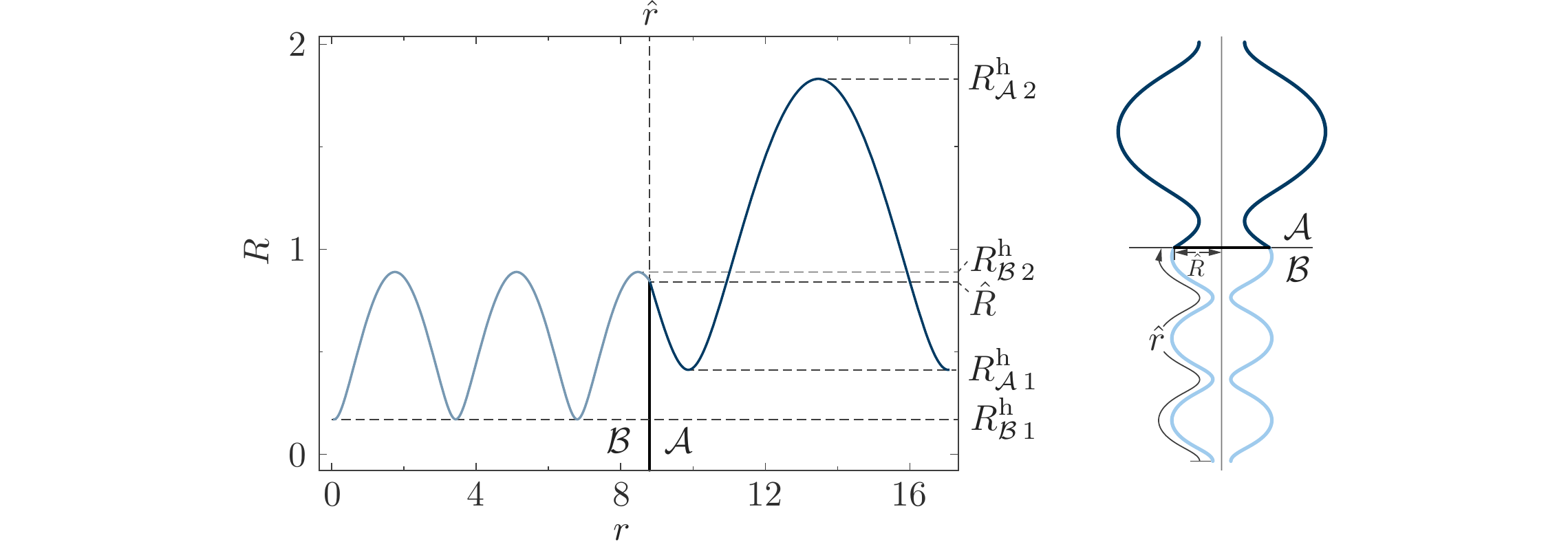}
  \caption{\small Left: Radius of curvature over the coordinate parametrizing a spatial geometry $r$ for multiple Schwarzschild-de Sitter spaces. In this example we have $N_{\cal B}=2$ disconnected regions in phase ${\cal B}$, and $N_{\cal A}=1$ disconnected region in phase ${\cal A}$. Right: The corresponding spatial geometry. \label{geometryexample}}
  \end{figure}

We consider transitions between a bound and an unbound turning point of a classical trajectory, where the spatial geometry satisfies $R^{\prime2}/L^2=A(R)$. At these points the domain wall radius of curvature is labeled by $\hat{R}_{1,2}$, where $\hat{R}_{1}<\hat{R}_2$, and we label the initial and final states by ${{\mathcal I}}$ and ${{\mathcal F}}$.
We call the horizons of the exterior spacetime $R^{\text{h}}_{{\cal A}\,1}<R^{\text{h}}_{{\cal A}\,2}$, where $A_{{\cal A}}(R^\text{h}_{{\cal A}\, 1})=A_{{\cal A}}(R^\text{h}_{{\cal A}\, 2})=0$, and similarly for the interior region ${\cal B}$. The contribution to the action from the spacetime regions between domain walls is  given by  (\ref{spacepart12}). At a classical turning point we have ${R^{\prime2}}=L^2 A_\alpha$, so the action simply becomes\footnote{The alert reader shall not be confused by signs when comparing to the literature: it is important to use consistent sign conventions. In this work we  use the convention $\text{arccosh}(-1)\equiv i \pi$, and $\arccos(-1)\equiv\pi$, while in parts of \cite{Fischler:1990pk} the sign convention appears to differ.}
  \bea\label{spaceinteg}
 i S_{\text{Space}}
 ={\eta_{\pi}\pi\over G}\int dR~ {R}\, {\Theta[-R^\prime] }\,,
  \eea
  where $\Theta$ is the Heaviside step function. Care has to be taken in picking the limits for the integral (\ref{spaceinteg}) as the integration proceeds along a path of increasing $r$. Using the Schwarzschild-(anti) de Sitter metric, and remembering that $R^\prime$ is negative between the outer and inner horizons $R^{\text{h}}_2$ and $R^{\text{h}}_1$ we can immediately evaluate the integral for the spatial contribution to the action away from the wall and find
  \bea\label{spaceactioneval}
{iS_{\text{Space},\hat{R}}}&=&{ \eta_{\pi} \pi\over 2G }\big[\Theta(-\hat{R}^\prime_{-})(\hat{R}^2-R^{\text{h}\,2}_{{\cal B} 2})+\Theta(-\hat{R}^\prime_{+})(R^{\text{h}\,2}_{{\cal A} 1}-\hat{R}^2)\nonumber\\&&+ N_{\cal A} (R^{\text{h}\,2}_{{\cal A}1}-R^{\text{h}\,2}_{{\cal A}2})+N_{\cal B} (R^{\text{h}\,2}_{{\cal B}1}- R^{\text{h}\,2}_{{\cal B}2})\big]\,,
\eea
where the integers $N_{{\cal A}/{\cal B}}$ count the number of spacetime regions in phase ${\cal A}$ or ${\cal B}$ that are disconnected from the domain wall, and vanishes for anti de Sitter spaces. This notation is illustrated by a specific example in Figure \ref{geometryexample}. 

To obtain the contribution to the action from non-trivial variations at the shells we evaluate the integral in (\ref{action2}) between the two classical turning points $\hat{R}_1$ and ${\hat{R}_2}$. While this is a hard integral in general, we can compute it analytically for the special case of vanishing asymptotic masses. Dropping the subscript of $\hat{R}_2$, and noting that $\hat{R}_1=0$, we find the action contribution at the  shell location
\be\label{shellactions}
iS_{\text{Shell}}=\begin{cases}\eta_{\pi}\pi \frac{ 2 \kappa  \left(H_{\cal A}^2-H_{\cal B}^2\right)+\kappa ^2 \hat{R} \left(H_{\cal B}^2+H_{\cal A}^2\right)+\hat{R} \left(H_{\cal B}^2-H_{\cal A}^2\right)^2}{8 G H_{\cal B}^2 H_{\cal A}^2 \kappa }\,,&\hat{R}^\prime_{-}<0\,,~\hat{R}^\prime_{+}<0\,,\\\eta_{\pi}\pi\frac{ (H_{\cal A}^2-H_{\cal B}^2)^2 \hat{R}+4\kappa H_{\cal A}^2  H_{\cal B}^2 \hat{R}^2+\kappa  (\kappa  \hat{R}-2)(H_{\cal A}^2+H_{\cal B}^2) }{8 G H_{\cal B}^2 H_{\cal A}^2 \kappa }
\,,&\hat{R}^\prime_{-}>0\,,~\hat{R}^\prime_{+}<0\,,\\\eta_{\pi}\pi\frac{  2 \kappa  \left(H_{\cal B}^2-H_{\cal A}^2\right)+\kappa ^2 \hat{R} \left(H_{\cal B}^2+H_{\cal A}^2\right)+\hat{R} \left(H_{\cal B}^2-H_{\cal A}^2\right)^2}{8 G H_{\cal B}^2 H_{\cal A}^2 \kappa }\,,&\hat{R}^\prime_{-}>0\,,~\hat{R}^\prime_{+}>0\,,\end{cases}
\ee
where the turning point corresponding to an unbound domain wall trajectory occurs at a radius of curvature
\be\label{radiuss}
\hat{R}^2=\frac{4\kappa^2}{{\left(H_{\cal A}^2-H_{\cal B}^2\right)^2+2 \kappa ^2 \left(H_{\cal A}^2+H_{\cal B}^2\right)+\kappa ^4}}\,.
\ee
As expected, the contribution to the action due to the shell is invariant under the exchange of insides and outside, which reverses the roles of $\cal A$ and $\cal B$, and switches the sign of $\hat{R}^\prime$. 

Finally, we are in a position to evaluate the probability for vacuum transitions. We add the contributions from the spatial, and shell parts of the action to find the exponent in (\ref{transitionprob}) that determines the leading contribution to the tunneling rate,
\be
\eta_\pi B=2i(S_{\text{Space},\,{\cal F}}+S_{\text{Shell},\, \cal{BA}}-S_{\text{Space},\,{\cal I}})\,.
\ee
With (\ref{spaceactioneval}) and (\ref{shellactions}) we then obtain the transition rate $\Gamma\sim e^{-|B|}$, where
\be
|{B}|=\pi\bigg| \hat{R}{ (H_{\cal A}^2 -H_{\cal B}^2)^2+(H_{\cal A}^2 +H_{\cal B}^2)\kappa^2\over 4 \kappa G H_{\cal A}^2H_{\cal B}^2 }+{N_{{\cal A},\,{\mathcal I}}-N_{{\cal A},\,{\mathcal F}}+{\text{sgn}(\hat{R}^\prime_{+})\over 2}\over GH_{\cal A}^2}+{N_{{\cal B},\,{\mathcal I}}-N_{{\cal B},\,{\mathcal F}}-{1\over 2}\over GH_{\cal B}^2}\bigg|\,,\label{tunnelingexp}
\ee
and we used $\eta_\pi=\text{sgn}(B)$ as prescribed by the matching conditions for the WKB tunneling wave function. The exponent (\ref{tunnelingexp}) is an important and beautiful result, so let us pause to appreciate some of its features. The first term in the tunneling exponent $B$ scales with the domain wall tension, while the latter two terms depend only on the Hubble scales in each of the phases. We wrote the tunneling rate such that the statistical nature of de Sitter space is manifest: the last two terms are proportional to the de Sitter  entropy ${\mathbb A}_{\text{dS}}/(4 G)$, where ${\mathbb A}_{\text{dS}}=4\pi H^{-2}$ is the horizon area.

Let us first recover the scenario considered by Coleman and de Luccia, where there exist no disconnected spacetimes\footnote{Some care has to be taken in determining the integers $N$ in this case. For example in the case of a true vacuum bubble where $R^\prime_+>0$ we have $N_{{\cal A},\,{\mathcal I}}=N_{{\cal A},\,{\mathcal F}}=1$ because the tunneling trajectory does not cross any horizon. In contrast, for a false vacuum bubble where $R^\prime_+<0$ we have $N_{{\cal A},\,{\mathcal I}}=1$ and $N_{{\cal A},\,{\mathcal F}}=0$ because the false vacuum bubble nucleates beyond the horizon of region ${\cal A}$. In both cases $N_{{\cal B},\,{\mathcal I}}=N_{{\cal B},\,{\mathcal F}}=0$. These observations follow immediately from (\ref{spaceinteg}).}
\be
N_{{\cal A},\,{\mathcal I}}-N_{{\cal A},\,{\mathcal F}}+{\text{sgn}(\hat{R}^\prime_{+})\over 2}={1\over 2}\,,~~N_{{\cal B},\,{\mathcal I}}-N_{{\cal B},\,{\mathcal F}}-{1\over 2}=-{1\over 2}\,.\label{cdlchoice}
\ee
The transition rate (\ref{tunnelingexp}) is then identical to the generalization of the CDL result to arbitrary extrinsic domain wall curvatures \cite{Parke:1982pm}. In Appendix \ref{app3} we rewrite the transition rate slightly to make this equivalence manifest. To gain some intuition for the probability of these transitions, consider the nucleation rate of bubbles occupied by phase ${\cal B}$ in a de Sitter vacuum ${\cal A}$, and compare this rate to that for the process where ${\cal A}$ and ${\cal B}$ are reversed. In the absence of wormholes, where (\ref{cdlchoice}) holds, this replacement always maintains the sign of $B$, so  the contribution to the decay rate at the wall cancels and we  find \cite{Lee:1987qc}
\be
{\Gamma|_{{\cal A}\rightarrow {\cal B}}\over \Gamma|_{{\cal B}\rightarrow {\cal A}}}=e^{{\cal S}_{\cal B}-{\cal S}_{\cal A}}\,,
\ee
where ${\cal S}_{{\cal A}/{\cal B}}={\pi /  GH_{{\cal A}/{\cal B}}^{2}}$ is the  entropy of a de Sitter space occupied by either phase. This means that semiclassical vacuum transitions appear to satisfy the principle of detailed balance in the absence of causally disconnected regions\footnote{Globally this is more complicated as the number of disconnected regions is measure dependent.}. Remember that the replacement ${\cal A}\leftrightarrow{\cal B}$ does {\it not} correspond to an exchange of  initial and final states, but the nucleation of a true vacuum bubble and the nucleation of a false vacuum bubble, respectively. Exchanging the initial and final spatial geometry ${\mathcal{I}}\leftrightarrow{\mathcal{F}}$ does not affect the transition probability when $\eta_{\pi}=\text{sgn}(B)$ in (\ref{transitionprob}). This is what one  expects in quantum mechanical tunneling process through a wide potential barrier in the WKB approximation because we consider the probability for a single incident wave packet to be transmitted through the barrier.

Let us now consider the nucleation of a false vacuum bubble in more detail. In the CDL scenario, where wormhole formation is prohibited, the transition rate approaches zero as the outside energy density $\rho_{\cal A}$ decreases. We can easily understand this by noting that the extrinsic curvature on the outside of the shell is negative, so the majority of the initial spacetime  disappears during the transition. The change in the action  increases with the inverse Hubble scale and diverges as ${{\rho}}_{\cal A}\rightarrow 0$, prohibiting the transition entirely. This corresponds to the well known result that Minkowski space cannot transition to a higher energy density. However, we can imagine a transition  that does not terminate the initial spacetime. Instead, consider a transition that maintains the entire initial region, nucleating a causally disconnected phase that contains the new vacuum. In this case the change in the total horizon area is independent of the initial Hubble scale. The spatial geometry of this transition is illustrated in the lower part of Figure \ref{tunnelcdl}. This case corresponds to the massless limit of the Farhi-Guth-Guven (FGG) process \cite{Farhi:1989yr}, for which we have $N_{{\cal B}\,,{\mathcal I}}=N_{{\cal B}\,,{\mathcal F}}=0$, $N_{{\cal A}\,,{\mathcal I}}=N_{{\cal A}\,,{\mathcal F}}=1$. Here the sign of $B$ becomes negative, so the tunneling wave function proposal indicates that $\eta_{\pi}=-1$. 
As expected, the transition rate remains finite in the limit of vanishing initial energy density
\be
\lim_{{{H}}_{\cal A}\rightarrow 0}B=-{\pi\over G} {H_{\cal B}^2+2\kappa^2\over (H_{\cal B}^2+\kappa^2)^2}\,.
\ee
The limit of vanishing domain wall tension  corresponds to a nucleation rate that is suppressed by the horizon area of the new de Sitter space, $\Gamma\sim e^{-\pi/GH_{\cal B}^2}$.

\subsection{An explicit tunneling trajectory}\label{explicittrajectory}

To gain some more intuition for the evolution of the spatial geometry during the vacuum transition process we now solve for an explicit, continuous tunneling trajectory that interpolates between the bound and unbound classical turning points $\hat{R}_1$ and $\hat{R}_2$. Again, we consider a single thin domain wall that connects two distinct Schwarzschild-de Sitter phases ${\cal B}$ (interior) and ${\cal A}$ (exterior). 
The transition is parametrized by the domain wall radius of curvature, $\hat{R}\in [\hat{R}_1,\hat{R}_2]$ and we denote the momentum evaluated at the domain wall with a hat, $\hat{\pi}_L\equiv \pi_L(\hat{R})$.   
\begin{figure}
  \centering
  \includegraphics[width=1\textwidth]{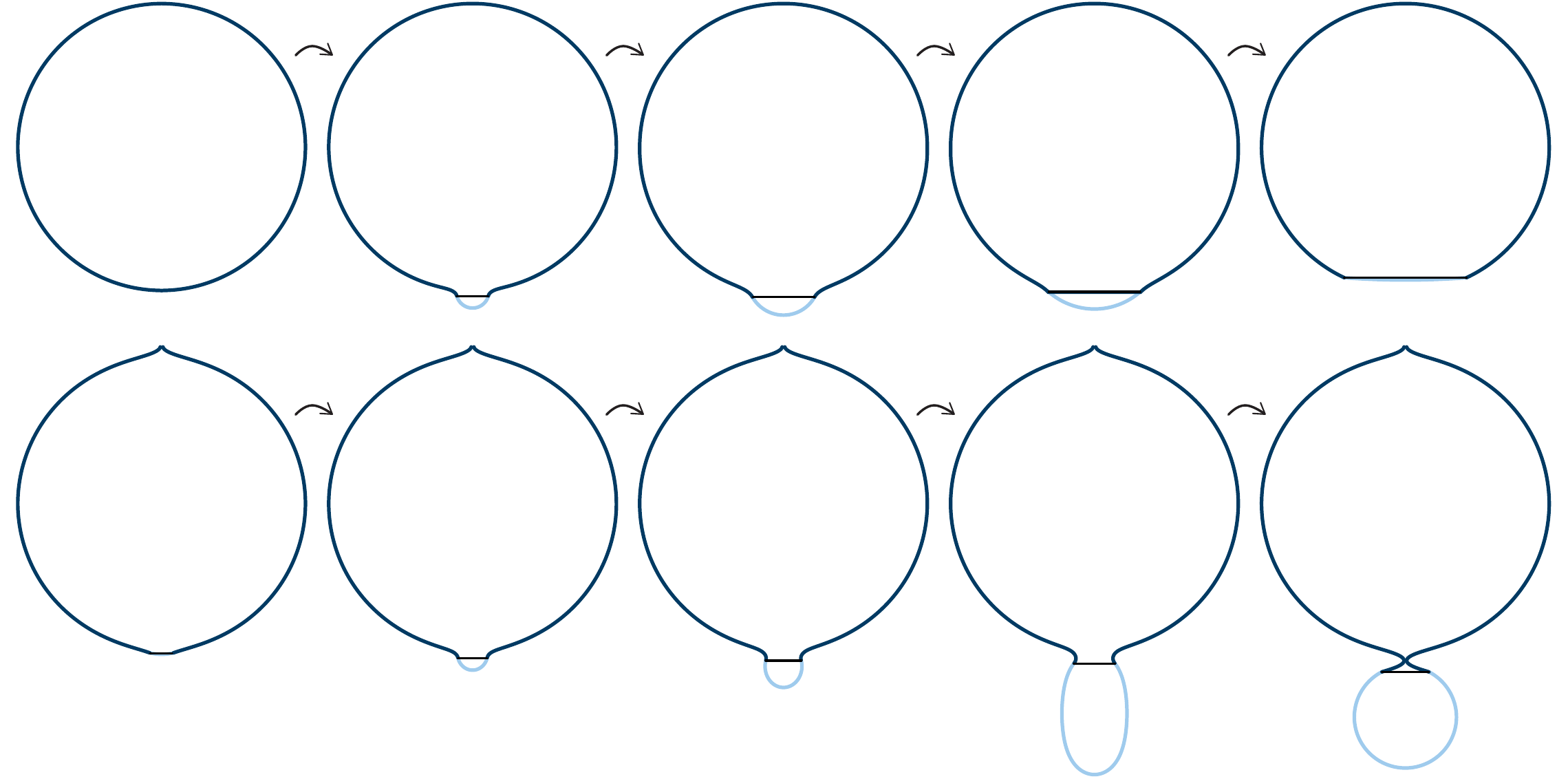}
  \caption{\small Illustration of continuous tunneling trajectories for the Coleman-de Luccia nucleation of a true vacuum bubble (upper row) and the Farhi-Guth-Guven nucleation of a false vacuum bubble (lower row). \label{tunnelcdl}}
\end{figure}

To obtain an explicit solution we demand $L=1$ and take the following ansatz for the momentum along the tunneling trajectory
\be\label{tunnelingtraj}
\pi_{L}(R)=\hat{\pi}_{L}f(R) \,,~~f(R)=\frac{(R^2-R^{\text{h}\,2}_1) (R^2-R^{\text{h}\,2}_2) }{(R^{\text{h}\,2}_1-\hat{R}^2) (R^{\text{h}\,2}_2-\hat{R}^2)}\left({\hat{R}\over R}\right)^{2\pm 2}\,,
\ee
where the positive sign in the exponent applies in the outer region ${\cal A}$, and the negative sign applies in the inner region ${\cal B}$. The horizons of each spacetime are labeled as $R^{\text{h}}_{1}<R^{\text{h}}_{2}$. The ansatz (\ref{tunnelingtraj}) is chosen such that the constraint equations (\ref{constraints}) and (\ref{rpsimplegauge}) are satisfied: the momentum vanishes at the horizons and takes on the correct value at the domain wall. 
We can now solve the constraint equation (\ref{constraints}) to obtain the three geometry specified by $R(r)$,
\be\label{threegeom}
R^{\prime 2}|_{{\cal A}/{\cal B}}={L^2 A_{{\cal A}/{\cal B}}(R)}+G^2L^2  \hat{\pi}^2_{L}f^2(R)|_{{\cal A}/{\cal B}}\,.
\ee
The differential equation (\ref{threegeom}) and the boundary conditions at the horizons fix the spatial geometry for any domain wall position $\hat{R}$ along the tunneling trajectory. The boundary conditions at horizons are necessary to specify whether the spatial geometry ends or continues behind the horizon, and if there are any domain walls in that patch.

Figure \ref{tunnelcdl} shows two specific trajectories for the quantum nucleation of true and false vacuum bubbles. During the nucleation of a true vacuum bubble the wall's extrinsic curvature does not switch sign\footnote{Some care has to be taken in evaluating the extrinsic curvature. Figure \ref{figurefull} might naively appear to imply that the extrinsic curvature does switch sign for the nucleation of a true vacuum bubble. However, that figure displays the case $0=M_{\cal A}<M_{\cal B}$, while the formation of a true vacuum bubble in the absence of any mass parameter corresponds to the opposite limit, $M_{\cal B}<M_{\cal A}\rightarrow 0$. In this limit (\ref{rpsds}) then yields the expected result of same sign extrinsic curvatures for the bound and unbound turning points.}, so the domain wall does not have to cross any horizons. Therefore, a tunneling trajectory parametrized by a monotonously increasing radius of curvature is continuous if the bubble nucleates within a causally connected region. This simply corresponds to the CDL transition. On the other hand, during the nucleation of a false vacuum bubble the wall's exterior extrinsic curvature does switch sign, which indicates that the wall crosses at least one horizon. The continuous trajectory corresponding to a monotonously increasing radius of curvature at the domain wall yields the formation of a false vacuum bubble through the Schwarzschild horizon. For the nucleation process shown in the lower panel of Figure \ref{tunnelcdl} the Schwarzschild horizon is crossed between the second and third steps. This simply corresponds to the FGG transition. The  CDL transition would correspond to false vacuum  nucleation through the Hubble horizon, and is not continuously parametrized by a monotonously increasing radius of curvature. Both the CDL and FGG processes of false vacuum nucleation are allowed at the semiclassical level, but correspond to different final geometries and have different transition rates.

\section{Gravitational Effects on Viable Vacua}\label{gravityeffects}
In the previous section we carefully reviewed thin wall vacuum transitions and we are finally in a position to address the question relevant to the landscape population: how are weakly coupled vacua populated in the presence of runaways? As discussed in \S\ref{domainwallsection}, there generically do not exist stable, thin domain walls separating vacua in the landscape. Instead, all domain walls connect to a runaway phase. In this hostile phase the simple four dimensional effective theory breaks down as the spacetime decompactifies, so we ought to revert to the higher dimensional theory to model this transition. To avoid having to face this much more difficult problem, in this work we model the runaway phase as a stable vacuum with vanishing energy density. It is conceivable that close to the domain wall the runaway instability is merely triggered, but the four dimensional description still provides a good approximation to local physics.
In this section we  address the question of whether and how stable vacuum transitions between de Sitter vacua ${\cal A}$ and ${\cal B}$ can occur if all domain walls connect to a runaway phase ${\cal C}$.
 This question was first discussed by Brown and Dahlen in \cite{Brown:2011ry}. They argued that even in the absence of a tunneling instanton any de Sitter vacuum transition will eventually occur due to the finite dimensionality of the Hilbert space. Even though our approach substantially differs from that work and we do not explicitly invoke the thermodynamic properties of de Sitter space, our results are compatible with and extend the results of \cite{Brown:2011ry}.
 
In order for a transition to persist at late times a horizon has to be crossed by the domain wall during the bubble nucleation process. Either the cosmological horizon of the original de Sitter phase or a wormhole horizon is traversed, which corresponds to the CDL or the FGG transition, respectively \cite{Coleman:1980aw,Farhi:1989yr,Aguirre:2005nt}. In this work we do not impose any constraints beyond obeying the Hamiltonian constraint equations, so we allow for both transitions. In the limit of a small initial vacuum energy density the FGG process is vastly more likely to occur because it preserves the initial spacetime, so we focus our discussion on that  solution. However, remember that the cosmological evolution after the transition is identical in both cases, so any constraints on the cosmology of the new phase apply regardless of which mechanism populates the vacua.

To summarize, in our model of the landscape ${\cal A/B}$ domain walls are prohibited, so any ${\cal A}\rightarrow{\cal B}$  transition between de Sitter vacua contains a double bubble with ${\cal B/C}$ and ${\cal C/A}$ domain walls, where ${\cal C}$ is an asymptotically flat region. We will consider the dynamics of these configurations and present the geometry and rate of a nucleation process that results in a stable vacuum transition between ${\cal A}$ and ${\cal B}$.

\subsection{The trouble with the  bubble}
\begin{figure}
  \centering
  \includegraphics[width=1\textwidth]{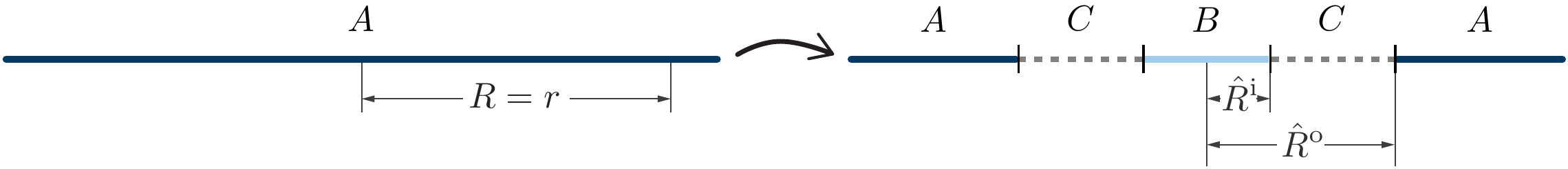}
  \caption{\small The spatial geometry of the formation of a double bubble in the absence of gravity. \label{nogravitybubble}}
  \end{figure}
As a warmup exercise, let us first consider a vacuum transition from ${\cal A}$ to ${\cal B}$ in the absence of dynamical gravity, $G\rightarrow 0$. In this limit we can write the metric (\ref{staticmetric}) simply as
\be
ds^2=-dT^2+dR^2+R^2 d\Omega^2_2\,,
\ee
or equivalently $L=1$ and $R=r$. The spatial configuration before and after the vacuum transition is shown in Figure \ref{nogravitybubble}. Initially the entire space is occupied by vacuum ${\cal A}$. Because of the absence of domain walls between ${\cal A}$ and ${\cal B}$, after the tunneling event there exists a region of phase ${\cal C}$ in between ${\cal A}$ and ${\cal B}$. We take the initial and final states to be in a pure vacuum configuration, such that the asymptotic masses vanish, $M_{\cal{A}}=M_{\cal{B}}=0$, while the intermediate phase ${\cal C}$ may experience a non-vanishing mass parameter. The vacuum energy densities of ${\cal A}$ and ${\cal B}$ are positive, but vanishes in ${\cal C}$. 
The constraint equation for the inner wall with radius of curvature $\hat{R}^\text{i}$ is given in (\ref{masseq}), which in the limit $G\rightarrow 0$ gives
\bea
M_{\cal C}&=&{4\pi\over 3}{{\rho}}_{\cal B}{\hat{R}^{\text{i}\,3}}+4\pi \sigma_{\text{i}} {\hat{R}^{\text{i}\,2}}\sqrt{1+\dot{\hat{R}}^{\text{i}\,2}}\,.
\eea
For an initially static configuration we immediately find the equation of motion for the inner domain wall 
\be
\ddot{\hat{R}}^\text{i}=-\left({{{\rho}}_{\cal B}\over \sigma_\text{i}}\sqrt{1+\dot{\hat{R}}^{\text{i}\,2}}+2{(1+\dot{\hat{R}}^{\text{i}\,2})\over \hat{R}^\text{i}}\right)<0\,,
\ee
inevitably leading to a collapse of the region in vacuum ${\cal B}$. As anticipated, in the absence of  horizons there does not exist a vacuum transition that would create a persistent region occupied by the new phase ${\cal B}$.

\subsection{No trouble with the double bubble}
\begin{figure}
  \centering
  \includegraphics[width=1\textwidth]{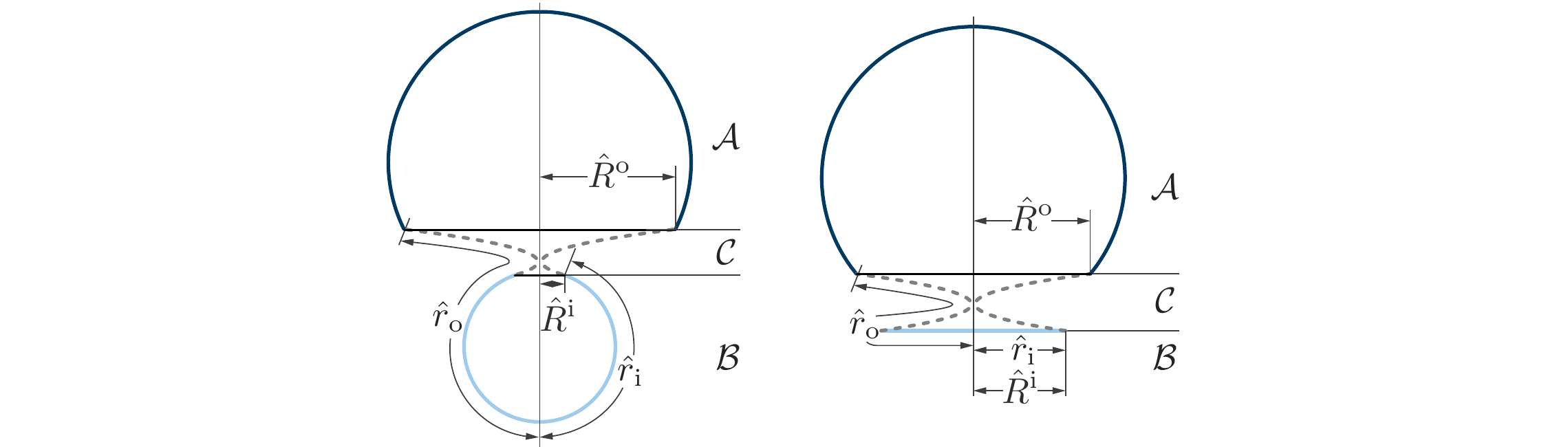}
  \caption{\small Two possible spatial configurations of a double bubble that leads to an unbound trajectory of the ${\cal B/ C}$ domain wall with small (left) and large (right) tension. \label{gravitybubble}}
\end{figure}
\begin{figure}
  \centering
  \includegraphics[width=1\textwidth]{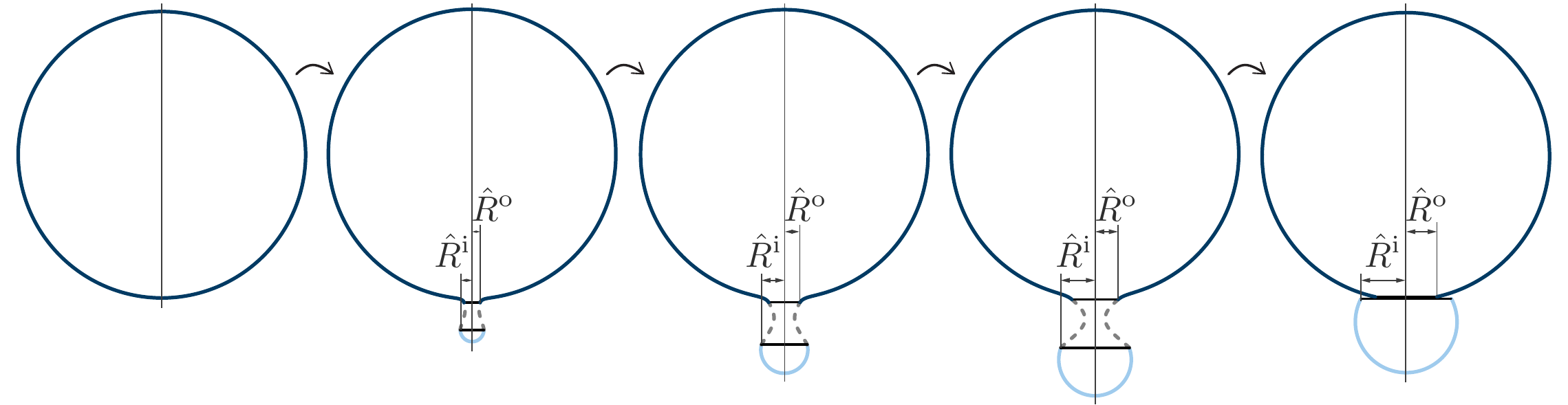}
  \caption{\small A continuous tunneling trajectory for the formation of a double bubble in the presence of gravity. The regions occupied by phases ${\cal A}$, ${\cal B}$ and ${\cal C}$ are illustrated with dark blue, light blue and dashed gray lines, respectively.\label{tunnelingdb}}
\end{figure}

Let us now turn to the transition between vacua ${\cal A}$ and ${\cal B}$ in the presence of gravity, but again in the absence of direct domain walls: only ${\cal C/A}$ and ${\cal B/C}$ domain walls exist. Gravity has a dramatic impact on the possible transition. In contrast to the situation without gravity, the initial and final spatial geometries are no longer fixed: depending on the energy densities, mass parameters, domain wall tensions and boundary conditions the spatial geometry changes. Two possible geometries after the transition are illustrated in Figure \ref{gravitybubble}. In the presence of gravity the radius of curvature of the inner shell can exceed the radius of the outer domain wall, while in the absence of gravity the corresponding instanton does not exist\footnote{This disappearance of the instanton without gravity was discussed in \cite{Brown:2011um}, but in the presence of gravity some of the instantons reappear.}.
 We  are interested in transitions for which at least the inner (${\cal B/C}$) domain wall grows without bound, leaving a part of the spacetime in vacuum ${\cal B}$. For simplicity we pick coordinates such that the transition occurs at $t=0$ where $L(t=0,r)=1$.  The domain wall dynamics can be read off from Figure \ref{figurefull2}. Note that the extrinsic curvature of the inner shell is negative in the exterior region, such that phase ${\cal C}$ contains a Schwarzschild horizon for any unbound solution. There are two qualitatively different unbound solutions. When the domain wall tension in Planck units dominates over the energy density in the interior, i.e. ${{\rho}}_{\cal B}<3\sigma_{\cal{BC}}^2/4\M^2$, there always exist unbound solutions where the extrinsic curvature changes sign across the shell. These domain walls expand due to their repulsive gravitational self-interaction and inflation inside phase ${\cal B}$ can be negligible. 
On the other hand, when the domain wall tension is small compared to the energy density, the  unbound domain walls have a negative extrinsic curvature on both sides of the domain wall. This is the familiar situation of a true vacuum bubble that has nucleated behind a wormhole horizon. In this case  the domain wall expands due to the different  energy density across the shell, and the gravitational self-interaction of the shell is negligible. Despite expanding without bound, the runaway phase ${\cal C}$ will never occupy all of the new de Sitter region because of the cosmological horizon in  vacuum ${\cal B}$. We can obtain the geometry along a tunneling trajectory by solving the constraint equation and junction conditions for the double bubble, as in \S\ref{explicittrajectory}. The continuous tunneling solution is shown in Figure \ref{tunnelingdb}.

\begin{figure}
  \centering
  \includegraphics[width=1\textwidth]{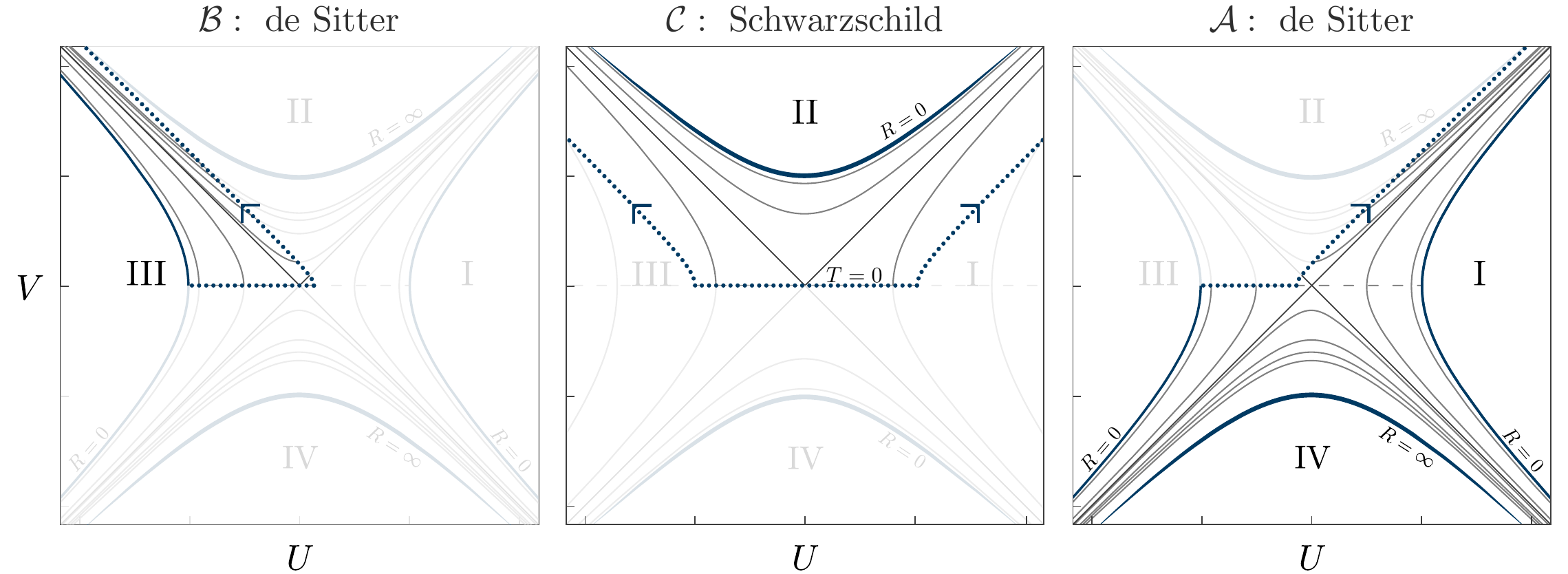}
  \caption{\small Illustration of the trajectories of a double bubble connecting de Sitter ($\cal B$), Schwarzschild ($\cal C$), and de Sitter ($\cal A$) patches. The bubble is nucleated at time $T=0$. Phase ${\cal B}$ contains a closed FRW cosmology, while ${\cal C}$ contains open FRW cosmologies on both sides of a wormhole \cite{ Coleman:1980aw,Farhi:1989yr,Bousso:2004tv}. \label{gravitybubbletime}}
\end{figure}

To give a concrete example in which we can easily understand both the initial geometry after tunneling and the subsequent classical dynamics, we consider solutions of vanishing mass parameter in each of the three regions. The junction conditions (\ref{rpsds}) at the domain walls  become
\bea
\dot{\hat{R}}^{\text{i}\,2}&=&{(\kappa_{\cal{BC}}^2+H_{\cal{B}}^2)^2\over 4\kappa_{\cal{BC}}^2} \hat{R}^{\text{i}\,2}-1\,,~~~\hat{R}^{\text{i}\,\prime}_{-}={\kappa_{\cal{BC}}^2-H_{\cal B}^2\over  2\kappa_{\cal{BC}}}\hat{R}^{\text{i}}\,,~~~\hat{R}^{\text{i}\,\prime}_{+}<0\,,\\
\dot{\hat{R}}^{\text{o}\,2}&=&{(\kappa_{\cal{CA}}^2+H_{\cal{A}}^2)^2\over 4\kappa_{\cal{CA}}^2} \hat{R}^{\text{o}\,2}-1\,,~~~\hat{R}^{\text{o}\,\prime}_{-}>0\,,~~~\hat{R}^{\text{o}\,\prime}_{+}={H_{\cal A}^2-\kappa_{\cal{CA}}^2\over  2\kappa_{\cal{CA}}}\hat{R}^\text{o}\,.
\eea
These are just the equations governing two expanding true vacuum bubbles, so none of the walls collapse into a singularity. For the inner domain wall we have the solution
\be\label{bcevolution}
\hat{R}^\text{i}={2\kappa_{\cal{BC}}\over \kappa_{\cal{BC}}^2+H_{\cal{B}}^2}\cosh\left({\kappa_{\cal{BC}}^2+H_{\cal{B}}^2\over 2\kappa_{\cal{BC}}}\,t\right)\,.
\ee
Note that after the tunneling event the spacetime in which the inner domain wall evolves is causally disconnected from original spacetime and the dynamics of the exterior domain wall are irrelevant. We show the classical domain wall evolution in each of the three regions in Figure \ref{gravitybubbletime}. The region in phase ${\cal C}$ contains two copies of open universes on opposite sides of a wormhole, while the region occupied by vacuum ${\cal B}$ contains a closed universe.

We now evaluate the tunneling rate to form a double bubble configuration from an initial de Sitter space occupied by vacuum ${\cal A}$. For simplicity we will only discuss the case valid in the limit of weak gravity, where $\kappa_{CA}<H_{\cal A}$, so the outside domain wall has positive extrinsic curvature, $\hat{R}^{o\prime}_{\pm}>0$. Again, the tunneling exponent is defined in (\ref{transitionprob}). The spatial contribution to the tunneling exponent is given with (\ref{spaceinteg}) as
\be
2i (S_{\text{Space},\, \mathcal{F}}-S_{\text{Space},\, \mathcal{I}})=\begin{cases}-{\eta_{\pi}\pi\over G}H_{\cal B}^{-2}\,,&\text{for}~\hat{R}_\pm^{\text{i}\,\prime}<0\\-{\eta_{\pi}\pi\over G}\hat{R}^{\text{i}\,2}\,,&\text{for}~\hat{R}_\pm^{\text{i}\,\prime}>0\end{cases}\,.
\ee
The contributions from the shell are given in (\ref{shellactions}), and we find the total tunneling exponent
\bea\label{doublebubbleb}
\eta_{\pi}B&=&|2i (S_{\text{Shell},\, \cal{CA}}+S_{\text{Shell},\, \cal{BC}}+S_{\text{Space},\, \mathcal{F}}-S_{\text{Space},\, \mathcal{I}})|\nonumber\\
&=&{\pi\over G}  \left|\frac{\kappa_{\cal{CA}}^4}{H_{\cal{A}}^2 \left(H_{\cal{A}}^2+\kappa_{\cal{CA}}^2\right)^2}-\frac{H_{\cal{B}}^2+2 \kappa_{\cal{BC}}^2}{\left(H_{\cal{B}}^2+\kappa_{\cal{BC}}^2\right)^2}\right|\,.
\eea
The corresponding transition rate $\Gamma\sim e^{-|B|}$ has a number of interesting properties. Let us begin with the limiting case of weak gravity, or small tension, where $\kappa$ is small compared to the Hubble scales involved. The tunneling rate is simply suppressed by the horizon area of the newly created de Sitter phase, $|B|=\pi /G H_{\cal B}^{2}$. Furthermore, the tunneling rate decreases as the Hubble scale of the initial vacuum ${\cal A}$ decreases. This feature is similar to the familiar nucleation of a true vacuum bubble: as the vacuum energy densities on both sides of the shell become equal the radius of the initial bubble increases without bound.  
Finally, there are two competing terms in the tunneling rate. When the two terms are competitive the tunneling rate becomes surprisingly large, i.e. $\Gamma\sim 1$. In the limit of a small outside domain wall tension, $\kappa_{CA}\ll H_{\cal A}$ the transition appears to be unsuppressed when
\bea
H_{\cal A}^3\sim \begin{cases}  \kappa_{\cal CA}^2\kappa_{\cal BC}/\sqrt{2}\,,~~~\kappa_{\cal BC}\gg H_{\cal B}\,,\\
\kappa_{\cal CA}^2 H_{\cal B}	\,,~~~~~~~~~\kappa_{\cal BC}\ll H_{\cal B}\,.
 \end{cases}
\eea
In either case the energy scale of the new phase is larger than the scales involved in the original vacuum and tunneling towards high energy configurations is favored. Naively, one might be concerned about unsuppressed transition rates, but we should remember that there may not be any metastable vacua at arbitrarily high energy that could be populated. Instead, the tunneling exponent (\ref{doublebubbleb}) implies that transitions among the highest (approximately) stable vacua are exponentially preferred. In this work we merely present one particularly simple thin wall process to illustrate the mechanism that gives rise to a stable configuration at late time, but other thick-wall transitions are possible.

The particular final geometry considered above is the leading transition channel for the population of a false vacuum with a high Hubble scale and a small outer domain wall tension. There are many other possible final states. For example, we could consider the nucleation of not just one, but $k$ disconnected de Sitter phases in vacuum ${\cal B}$. For small domain wall tensions the nucleation rate is roughly given by
\be{\label{kdsregions}}
\Gamma\sim e^{-\pi k /GH_{\cal B}^{2}}\,,
\ee
which greatly suppresses the nucleation of disconnected universes. For example, in the case of a metastable vacuum with almost Planckian energy density $\rho_{\cal B}\sim 0.1\times \M^4$ the vacuum nucleation rate is roughly $\Gamma \sim e^{-2000\times k}$. Note that this result is crucially dependent on the sign choice we made for $\eta_{\pi}$. Abandoning the tunneling wave function and choosing $\eta_{\pi}=+1$ instead would have given a divergent rate as the number of disconnected universes grows \cite{Fischler:1990pk}.

\subsection{Wormholes in quantum gravity}
Wormholes have been the focus of intense research for many decades \cite{ArkaniHamed:2007js,Hawking:1987mz,Hawking:1988ae,Coleman:1988cy,Giddings:1988cx,Coleman:1988tj,Coleman:1990tz,Rey:1998yx,Maldacena:2004rf,Giddings:1989bq,Bergshoeff:2004fq,Rubakov:1996cn,Kim:1997dm,Cadoni:1994av,Harlow:2015lma,Montero:2015ofa,Bachlechner:2015qja}, and yet their role in nature is far from clear. 
We saw in the previous subsection that in the absence of stable domain walls between cosmological vacua bubble nucleation occurs across an event horizon that  stabilizes the transition. An example of such a transition is the FGG process, which nucleates a wormhole geometry. Much effort has been spent on studying whether such an event is admissible in quantum gravity, but no definite conclusion has been reached.
Rather than providing a comprehensive literature review, we refer the interested reader to a number of relevant works on the subject, see  \cite{Banks:2002nm,Banks:2007ei,Aguirre:2005nt,Freivogel:2005qh,Aguirre:2005xs,Bousso:2004tv,Banks:2003es,Banks:2004xh,Banks:2012hx,Banks:2010tj}.

In discussions that employ a semiclassical approximation such that ambiguities about the quantization of gravity are irrelevant all classical solutions to the Hamilton-Jacobi are treated on equal footing. Much care has to be taken when attempting to constrain or disregard some of the classical solutions based on presumed knowledge  about 	quantum gravity. The Einstein-Hilbert action is well understood, while the full quantum mechanical description of wormholes remains elusive. Vague arguments and beliefs about how quantum gravity ought to behave may be misleading.

\section{Inflation in the Landscape}\label{inflatinsection}
In the previous section we discussed a tunneling geometry that facilitates vacuum transitions between two de Sitter vacua in the absence of a direct domain wall. While we illustrated one particular trajectory that induces  stable transitions, there may be many other ways to populate the landscape, such as dynamical and thick-wall solutions that are beyond the reach of our analysis. However, regardless of the precise geometry  after the transition, the new phase is causally protected from domain wall collapse only if the runaway phase remains outside the Hubble horizon. We now  consider the implications for the cosmological evolution in the new vacuum. This discussion is  independent of the explicit tunneling trajectory and  applies for general transitions with thin domain walls. The basic idea is simple: the new phase is causally protected from collapse of the shell due to the cosmological horizon, but if the  evolution allows the domain wall to enter the Hubble sphere of a late time observer the new universe is at risk of extinction. In an evolving cosmology this  means that the initial inflationary phase should last long enough to remove the domain wall from the late time horizon.

\subsection{Inflation expels runaways}
We now turn to the cosmology after the tunneling event. To be specific, we  discuss an FRW cosmology that undergoes a finite period of inflation, followed by radiation domination, and finally settles into energy domination at late times. Shortly after the tunneling event the relevant spacetime is divided into two regions: the new phase in vacuum ${\cal B}$ is described by a closed FRW universe and is separated from the runaway phase ${\cal C}$ by an expanding domain wall. The region occupied by the runaway phase with vanishing energy density is described by an open FRW cosmology with a (very small) black hole. We are most curious about the cosmology of the newly created universe in vacuum ${\cal B}$. So far we only considered a stationary, energy dominated interior with a constant equation of state $w=p/\rho=-1$. We now relax this condition, and consider an evolution that maintains the initial inflationary Hubble scale $H_{\text{Inf}}$ only for a finite amount of expansion and is  followed by reheating, radiation domination and finally a classical evolution towards a (potentially much lower) late time Hubble scale $H_{\text{Late}}$. Given sufficient expansion to overcome the initial curvature domination, we can approximate the metric by a flat FRW cosmology. The causal structure is most obvious when expressing the metric in terms of conformal time $\tau$, such that the metric takes the form
\be
ds^2=a^2(-d\tau^2+dR^2+R^2d\Omega^2_2)\,.
\ee
\begin{figure}
  \centering
  \includegraphics[width=1\textwidth]{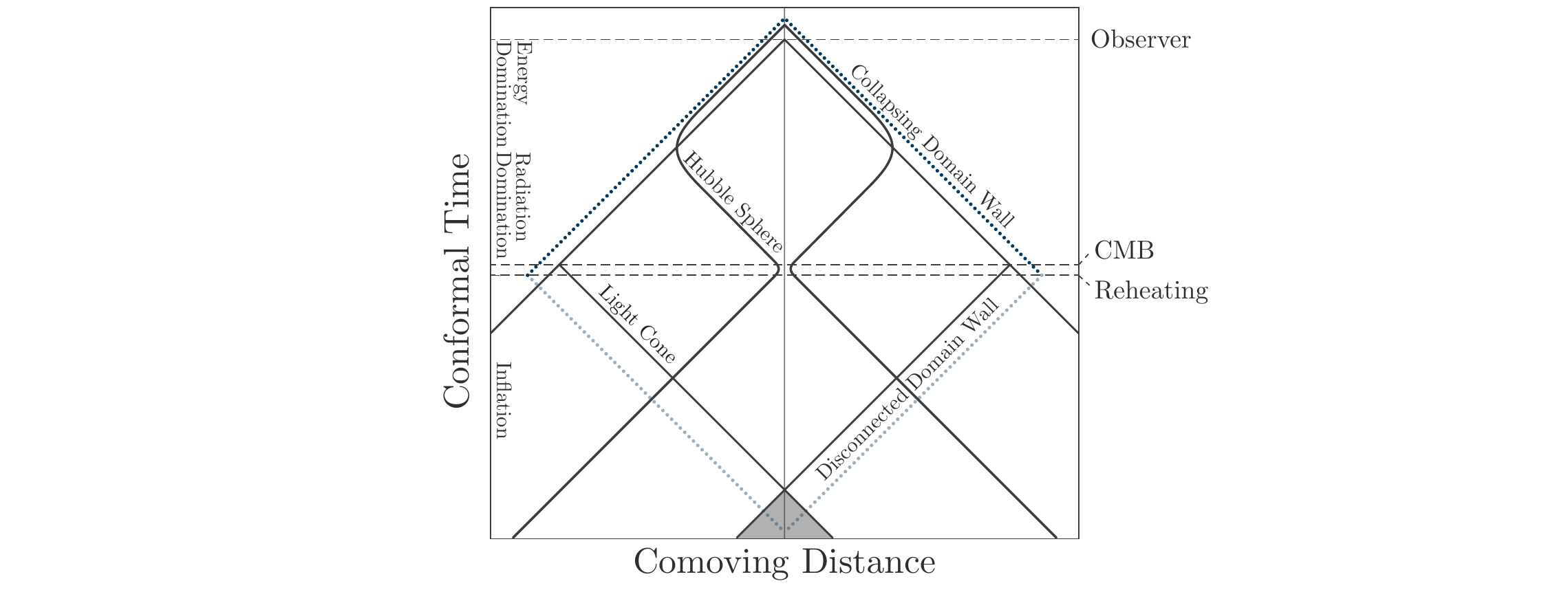}
  \caption{\small Conformal time and comoving distance for a universe undergoing inflation, radiation domination and a late phase of energy domination \cite{Baumann:2014nda}. During inflation a runaway domain wall expands in a causally disconnected region. After reheating, the domain wall might collapse and re-enter the horizon unless all observable past light-cones were in causal contact. \label{conformaltime}}
\end{figure}

During inflation and late time energy domination the comoving Hubble sphere shrinks with conformal time, $(aH)^{-1}_{\text{Inf}}\approx-\tau+\text{const}$, while during radiation domination the horizon expands and allows modes to enter, $(aH)^{-1}_{\text{Rad}}\approx\tau+\text{const}$.
After the nucleation event the domain wall separating the new cosmology from the runaway phase initially accelerates outwards along an approximately null trajectory in a causally disconnected region of de Sitter space. For domain wall tensions that are small compared to the inflationary Hubble scale we have $R^\prime_-<0$, so the domain wall is hidden outside the Hubble horizon as in the left part of Figure \ref{gravitybubble}. At the end of inflation the equation of state changes. The precise evolution of the bubble will depend on the dynamics during reheating and radiation domination, but it is possible that after inflation the energy density  ${\cal B}$ is small compared to the tension $\kappa$, so the spatial geometry corresponds to the right part of Figure \ref{gravitybubble}. This scenario appears likely if we demand a vacuum energy density small enough to allow for galaxy formation. In this case $R^\prime_->0$ and the domain wall is at risk of re-collapsing and terminating newly created universe. 
However, if the late time cosmology is dominated by a positive energy density, it is possible that even a collapsing domain domain wall never re-enters the horizon. This will be the case if the bubble radius exceeds the late time Hubble scale. The sufficient condition for the existence of a region of spacetime to survive indefinitely can be seen in Figure \ref{conformaltime}, which shows the evolution of comoving distance with conformal time. The condition for the domain wall to remain out of causal contact with a late time is precisely the requirement that all null geodesics originating from the time of reheating have overlapping past light cones, and is slightly stronger than the requirement of super-horizon correlations in the CMB. We find a rough lower bound for the required number of efolds as
\be\label{mininfl}
N_e\gtrsim{1\over 2}\log\left( {H_{\text{Inf}}\over H_{\text{Late}}}\right)\,,
\ee
which corresponds to about $65$ efolds of expansion, depending on the scale of inflation\footnote{We thank Matthew Kleban for discussion on this point.}. This amount of inflation leaves the universe in a surprisingly flat and isotropic state.

Even though inflation clearly can exile a collapsing domain wall from the late time cosmology and save the universe from its demise, it is not obvious that this is a necessary condition. Remember that when the domain wall tension is large, or gravity sufficiently strong, the self interaction bubble wall is repulsive and leads to an expanding domain wall in an empty universe.  There may be concerns associated with strong gravitational interactions\footnote{It may be a curious coincidence that in a simple axion model the low-tension constraint $\kappa<H$ typically coincides with a naive formulation of the weak gravity conjecture $f\lesssim\M$, where $f$ is the axion decay constant and we assumed a large instanton action, $S>1$.} \cite{Vafa:2005ui,ArkaniHamed:2006dz,Rudelius:2014wla,Cheung:2014vva,Heidenreich:2015wga,Bachlechner:2015qja,Rudelius:2015xta,Hebecker:2015rya,Ibanez:2015fcv,Heidenreich:2016aqi}. If the transitions corresponding to strong gravity are indeed prohibited, the only way to remove the domain wall from a late time cosmology is via a period of inflation.

 The domain wall dynamics after reheating will depend on the details of the cosmological evolution, so even though this may not seem  likely, there could be some non-trivial evolution that halts domain wall collapse   within the cosmological horizon.  A detailed study of the cosmological evolution after reheating is beyond the scope of this work \cite{Tanahashi:2014sma}.

\subsection{Cosmology in the landscape}
Finally, we are in a position to speculate about the cosmological evolution in a landscape when both our  assumptions about fundamental physics are met: there exist no direct domain walls between de Sitter vacua, and the tunneling wave function provides a good approximation to vacuum transition rates.

Let us consider an initial state with an asymptotically flat background geometry in a phase  ${\cal C}$. This may be a stable ground state of the four dimensional effective theory or the decompactified runaway phase. Gravitational effects stabilize the nucleation of a false vacuum bubble containing the de Sitter vacuum ${\cal A}$, separated by a single domain wall from the initial phase ${\cal C}$. This bubble nucleates across a wormhole horizon as in the FGG process, and the nucleation rate is given by (\ref{tunnelingexp}). In the limit of weak gravity the transition rate is suppressed by the horizon area of the new phase, so transitions to high energy vacua are exponentially favored\footnote{Relaxing our assumption of employing the tunneling wave function and picking $\eta_{\pi}=+1$ instead would give a divergent nucleation rate in the semiclassical approximation. We do not discuss this case further.},
\be
\Gamma\sim e^{-{\pi / GH^{2}_{\text{Inf}} }}\,.
\ee
The tunneling process gives rise to a closed FRW cosmology in vacuum ${\cal A}$ that is causally disconnected from most of the original spacetime. The runaway phase ${\cal C}$ remains mostly undisturbed. 

Depending on the cosmological evolution in the new phase ${\cal A}$, the bubble may collapse, drive eternal inflation, or result in a viable late time cosmology. Because of the absence of direct domain walls to other de Sitter vacua with lower energy density, any vacuum transition inside the de Sitter phase ${\cal A}$ will trigger an expanding true vacuum bubble containing the runaway phase ${\cal C}$. However, the nucleation of a double bubble can induce a vacuum transition to a different de Sitter vacuum ${\cal B}$. The rate for this process to occur is given in (\ref{doublebubbleb}). Again, tunneling towards high energy states are favored. If the energy density of the domain wall is set approximately by the same scale as the initial vacuum energy density inside the cosmology, the domain wall tension in Planck units is negligible and the domain wall is only guaranteed to expand if an extended period of inflation takes place in the new vacuum ${\cal B}$. Again, there are two interesting scenarios: either the new phase is classically stable and leads to more eternal inflation, or the equation of state changes and relaxes to a lower Hubble scale $H_{\text{Late}}$ that allows for galaxy formation. The latter case is safe from domain wall collapse if all observable modes from the time of reheating have overlapping past light-cones. This requirement translates to a minimum amount of inflationary expansion set by (\ref{mininfl}), and is sufficient to create a surprisingly flat and isotropic universe for late time observers to occupy.

\section{Conclusions}\label{conclusionss}
We considered thin wall vacuum transitions in the absence of domain walls interpolating between metastable de Sitter vacua, allowing only for domain walls between the de Sitter regions and a runaway phase with vanishing vacuum energy density. This setup is motivated by the Dine-Seiberg problem in weakly coupled compactifications of string theory.
 Despite the instability of domain walls, there exist vacuum transitions between de Sitter vacua and the landscape is populated by quantum tunneling. In the weak gravity limit the leading transitions are mediated via a double bubble configuration that contains a wormhole and is illustrated in Figure \ref{gravitybubble}. For low domain wall tensions the nucleation rate is suppressed by the de Sitter horizon area of the nucleated phase, favoring transitions towards high Hubble scales. These transitions can be interpreted as small, local fluctuations to low entropy states that are subsequently frozen by the formation of a Hubble horizon. The new cosmology is protected from  domain wall collapse as long as the shell does not enter the horizon of an observer, which imposes a severe constraint on the cosmological evolution. The domain wall  can re-enter the  horizon unless a sufficient amount of expansion has taken place to permanently exile the shell from a late time horizon $H^{-1}_\text{Late}$. Demanding that the cosmological phase is causally protected from a possible re-collapse of the domain wall gives a lower bound on the number of efolds of inflation,
\be\label{mininfl2}
N_e\gtrsim{1\over 2}\log\left( {H_{\text{Inf}}\over H_{\text{Late}}}\right)\,.
\ee
This lower bound on the inflationary expansion guarantees overlapping past light cones for all observable modes and results in an isotropic and flat universe.

In this work we employed two well motivated assumptions about fundamental physics and arrived at surprisingly strong implications for the cosmological evolution in a landscape. We saw that generic instabilities of weakly coupled string theory vacua require an extended period of cosmic inflation to stabilize vacuum decay, and that the tunneling wave function approach to transition rates exponentially favors high inflationary scales. Both of these effects provide potentially much stronger selection effects than some of the previously assumed measures in theory space, such as a polynomial suppression in the number of efolds that stems from the requirement of small slow roll parameters in a random potential. Therefore, a detailed understanding of moduli stabilization and vacuum transitions in theories of quantum gravity are important to further our understanding, and ultimately make predictions, of inflationary parameters in the landscape. 

Our results point towards framework to pursue the generation of inflationary initial conditions in  flux compactifications and its explicit realization in a well controlled compactification of string theory is an important problem for the future.

\section*{Acknowledgements}
I  would like to thank Adam Brown, Kate Eckerle, Frederik Denef, Matthew Johnson, Matthew Kleban,  Liam McAllister, Ruben Monten, Henry Tye, Eve Vavagiakis, Alexander Vilenkin, Erick Weinberg, Timm Wrase and Claire Zukowski for useful discussions. 
This work was supported by the U.S. Department of Energy (DOE) under under grant no. DE-SC0011941. This research was supported in part by Perimeter Institute for Theoretical Physics. Research at Perimeter Institute is supported by the Government of Canada through the Department of Innovation, Science and Economic Development and by the Province of Ontario through the Ministry of Research, Innovation and Science.

\appendix
\section{First-Order Action of Domain Walls}\label{app1}
The results in this work are derived from the first-order action of a spherically symmetric gravitational configuration. Even though the relevant action has appeared several times in the literature, a number of signs are inconsistent in some of the references. Since our work crucially relies on the accuracy of the relative signs in the action, we now fill in some of the details omitted in the main text. We will mostly follow the references \cite{Fischler:1989se,Fischler:1990pk,Kraus:1994by}.

The dynamical terms in the action (\ref{ehcation}) can be written in terms of the variables appearing in the metric (\ref{sphericalmetric}) as \footnote{Again, partial derivatives with respect to $r$ and $t$ are denoted by primes and dots, respectively.} \cite{Fischler:1990pk}
\bea
S^\text{Gravity}={1\over 2 G}\int dr dt~\bigg(&& {2\over N^t}(N^rLR)^{\prime} (\dot{R}-N^r R^\prime )-{2\over N^t}\partial_t (L R)(\dot{R}-N^r R^\prime)  \nonumber\\
&&+{2\over L}(N^t R)^\prime R^\prime+ {N^t\over L}(L^2-R^{\prime2}) +{L\over N^t}(\dot{R}-N^r R^\prime)^2\bigg)\,,
\eea
which can be written in terms of canonical variables as
\be
S^\text{Gravity}=\int dr dt~\left(\pi_L\dot{L}+\pi_R \dot{R}-N^t {\cal H}_t^{\text{Gravity}}-N^r {\cal H}_r^{\text{Gravity}}\right)\,,
\ee
where
\be
\pi_L={N^r R^\prime-\dot{R}\over G N^t}R\,,~~
\pi_R={(N^r L R)^\prime - \partial_t (L R)\over GN^t },,
\ee
and the Hamiltonian density is given by
\bea
{\cal H}_t^{\text{Gravity}}&=&{G L \pi_L^2\over 2 R^2}-{G\over R} \pi_L\pi_R+{1\over2 G}\left[\left({2 R R^\prime\over L}\right)^\prime -{R^{\prime2}\over L}-L\right]\,,\nonumber\\
{\cal H}_r^{\text{Gravity}}&=&R^\prime \pi_R-L\pi_L^\prime\,.
\eea
We can also solve for the velocities $\dot{R}$ and $\dot{L}$,
\be
\dot{R}=-{G N^t\pi_L\over R}+N^r R^\prime\,,~~~\dot{L}=-{G N^t \pi_R\over R}+{G L N^t\pi_L\over R^2}+(N^r L)^\prime\,.
\ee
It remains to write the matter contributions in terms of canonical variables. For simplicity we consider a matter energy momentum tensor that originates from a static scalar field theory with thin domain walls, such that the domain wall tension equals the domain wall energy density. We then have the matter action
\bea
\int dx^4~\sqrt{g}{\cal L}_m&=&-4\pi \int dtdr\sqrt{g} R^2\left(\rho(r)+\sum_{\alpha=1}^N \sigma_\alpha \delta(r-\hat{r}_\alpha)\right)\\
&=&-4\pi \int dtdr L N^t R^2\rho(r) -4\pi\int \sum_{\alpha=1}^{N-1} \sigma_\alpha \hat{R}^2 \sqrt{-\hat{g}_{\mu \nu} dx^\mu dx^\nu }\nonumber\\
&=&-\int dt dr~ N^t {\cal H}_t^{\rho}+ \sum_{\alpha=1}^{N-1} \int dt ~\hat{p}_\alpha\dot{\hat{r}}_\alpha-\int dtdr~ (N^t {\cal H}_{t,\alpha}^\text{Shell} +N^r {\cal H}_{r,\alpha}^\text{Shell})\nonumber\,,
\eea
and we defined
\bea
m_\alpha=4\pi \sigma_\alpha \hat{R}^2_\alpha\,&,&~~
\hat{p}_\alpha={m_\alpha\hat{L}^2_\alpha (\hat{N}^r_\alpha+\dot{\hat{r}}_\alpha)\over \sqrt{\hat{N}^{t2}_\alpha-\hat{L}_\alpha^2(\dot{\hat{r}}_\alpha+\hat{N}_\alpha^r)^2}}\,\\
{\cal H}_{t,\alpha}^\text{Shell}= \delta(r-\hat{r}_\alpha) \sqrt{{\hat{p}_\alpha^2\over L^2}+m_\alpha^2}\,&,&~~
{\cal H}_{r,\alpha}^\text{Shell}=-\delta (r-\hat{r}_\alpha) \hat{p}_\alpha\,,~~
{\cal H}_t^{\rho}=4\pi L R^2 \rho(r)\,.\nonumber 
\eea
Let's also define a matter Hamiltonian density that includes the contribution from a vacuum energy density and the domain wall tension,
\bea
{\cal H}^{\text{Matter}}_t={\cal H}_t^{\rho}+\sum_{\alpha=1}^{N-1}{\cal H}_{t,\alpha}^\text{Shell}\,,~~
{\cal H}^{\text{Matter}}_r=\sum_{\alpha=1}^{N-1}{\cal H}_{r,\alpha}^\text{Shell}\,.
\eea
Combing the above expressions, we can write the  action as
\be\label{action2app}
S= \sum_{\alpha=1}^{N-1} \int dt ~\hat{p}_\alpha\dot{\hat{r}}_\alpha+\int dtdr~\left( \pi_L\dot{L}+\pi_R \dot{R}-N^t {\cal H}_t-N^r {\cal H}_r\right)\,,
\ee
where the total Hamiltonian density is given by
\be
{\cal H}_{t,r}={\cal H}_{t,r}^\text{Gravity}+{\cal H}_{t,r}^\text{Matter}\,.
\ee
The Hamiltonian densities in full are given by 
\bea
&&{\cal H}_t={G L \pi_L^2\over 2 R^2}-{G\over R} \pi_L\pi_R+{\left({2 R R^\prime\over L}\right)^\prime -{R^{\prime2}\over L}-L\over2 G}+4\pi L R^2 \rho(r)+\sum_{\alpha=1}^{N-1}\delta(r-\hat{r}_\alpha) \sqrt{{\hat{p}_\alpha^2\over L^2}+m_\alpha^2}\nonumber,\\
&&{\cal H}_r=R^\prime \pi_R-L\pi_L^\prime-\sum_{\alpha=1}^{N-1}\delta (r-\hat{r}_\alpha) \hat{p}_\alpha\,,\label{junction2a}
\eea
Note that the action (\ref{action2app}) contains second derivatives. Restricting to asymptotically flat solutions, where $\lim_{r\rightarrow \infty} N^t= 1$ and $\lim_{r\rightarrow \infty} N^r= 0$, the second derivative terms are precisely canceled by including a non-dynamical term $-\int dt M_{\text{ADM}}$ \cite{Kraus:1994by,Regge:1974zd}. In the present case we only consider solutions with constant asymptotic mass and therefore the ADM term is irrelevant.

Finally, we turn to the question relevant for quantum tunneling, where we need to evaluate the action for an arbitrary shell trajectory. This shell trajectory will parametrize the tunneling event. The variation of the action is given by
\be
dS=\sum_{\alpha=1}^{N-1} \hat{p}_\alpha \delta\hat{r}_\alpha+\int dr~\left( \pi_L\delta L+\pi_R \delta R\right)\,.
\ee
To integrate this expression we keep $\delta R=\delta\hat{r}_\alpha=\delta\hat{L}_\alpha=0$ and vary $L$ away from the shells to satisfy the constraints. This gives the following  contribution to the action \cite{Fischler:1990pk},
\bea
S_{\text{Space}}&=&\sum_{\alpha=1}^N\int_{\hat{r}_{\alpha-1}+\epsilon}^{\hat{r}_{\alpha}-\epsilon} dr\int_{\pi_L=0}^L(\delta L \pi_L+\delta R \pi_R)\nonumber\\
&=&\sum_{\alpha=1}^N\left(\int_{\hat{r}_{\alpha-1}+\epsilon}^{\hat{r}_{\alpha}-\epsilon} dr\int_{\pi_L=0}^L\delta L~{R\eta_\pi\over G}\sqrt{{R^{\prime2}\over L^2}-A_\alpha}\right)\nonumber\\
&=&\sum_{\alpha=1}^N\int_{\hat{r}_{\alpha-1}+\epsilon}^{\hat{r}_{\alpha}-\epsilon} dr~{R \eta_{\pi}\over G}\left(\sqrt{{R^{\prime2}}-L^2 A_\alpha} - {R^\prime }\log\left[{{R^\prime }-\sqrt{{R^{\prime2}}-L^2 A_\alpha}\over L \sqrt{A_\alpha} }\right]\right)\,.
\eea
Note that $R^\prime$ is discontinuous at the location of the shells. Considering an arbitrary variation of $R$, the action receives an additional contribution at the location of shells,
\be
\delta S_{\text{Shell},\,\alpha}= \delta \hat{R}^\prime\left({\partial S_{\space}\over \partial R^\prime}\bigg|_{r=\hat{r}_{\alpha}-\epsilon} - {\partial S_{\space}\over \partial R^\prime}\bigg|_{r=\hat{r}_{\alpha}+\epsilon}\right)\,.
\ee
We need to subtract derivatives with respect to $R^\prime$ at the location of the shell to ensure that the full action satisfies ${\delta S/\delta R}=\pi_R$ and ${\delta S/ \delta L}=\pi_L$. This gives
\bea
S_{\text{Shell},\,\alpha}&=&\int dt\,\hat{p}_\alpha \dot{\hat{r}}_\alpha-  \hat{R}^\prime\left({\partial S_{\space}\over \partial R^\prime}\bigg|_{r=\hat{r}_{\alpha}-\epsilon} - {\partial S_{\space}\over \partial R^\prime}\bigg|_{r=\hat{r}_{\alpha}+\epsilon}\right)\\
&=&\int dt \,\hat{p}_\alpha\dot{\hat{r}}_\alpha+\int d\hat{R}_\alpha\,{\hat{R}_\alpha\eta_{\pi}\over G}\log\left( {{\hat{R}^\prime}_{\alpha,+}-\sqrt{{\hat{R}^{\prime2}}_{\alpha,+}-\hat{L}^2_\alpha\hat{A}_{\alpha+1}}\over {\hat{R}^\prime}_{\alpha,-}-\sqrt{{\hat{R}^{\prime2}}_{\alpha,-}-\hat{L}^2_\alpha\hat{A}_{\alpha}} }\sqrt{\hat{A}_\alpha\over \hat{A}_{\alpha+1}}\right),\nonumber
\eea
where $\hat{A}_{\alpha+1}\equiv \hat{A}_{\alpha+1}(\hat{r}_\alpha)$ and the argument of $\hat{A}_{\alpha+1}$ is implicit. Collecting all the terms then gives the full action quoted in (\ref{action2}) in the main text,
\be
S=S_{\text{Space}}+\sum_{\alpha=1}^{N-1}S_{\text{Shell},\,\alpha}\,.
\ee

\subsection{The junction conditions}
The action does not contain a kinetic term for the shift and lapse functions, we therefore have the Hamiltonian constraints 
\be
{\cal H}_{t}={\cal H}_{r}=0\,.
\ee
We  obtain the required $2(N-1)$ constraint equations that fix the classical dynamics of the system by considering a linear combination of these constraints at the location at each shell. Following \cite{Fischler:1990pk,Kraus:1994by}, consider the following constraint
\be
0={R^\prime\over L}{\cal H}_{t}+{\pi_L\over R L}  {\cal H}_{r}=-{ M}^\prime +\sum_{\alpha=1}^{N-1}{\hat{R}^\prime_\alpha\over \hat{L}_\alpha}{\cal H}_{t,\alpha}^\text{Shell}+\sum_{\alpha=1}^{N-1}{G\hat{\pi}_{L,\alpha}\over \hat{R}_\alpha \hat{L}_\alpha} {\cal H}_{r,\alpha}^\text{Shell}\,,
\ee
where
\be
{M}={G\pi_L^2\over 2R} +{R\over 2G} \left(1-{R^{\prime 2}\over L^2}\right)-{4\pi} \int dR~R^2 \rho(r)\,.
\ee
In a static spacetime, ${ M}$ can be interpreted as the asymptotic mass parameter. For constant energy densities within the domains $\alpha$, we can evaluate the derivative ${ M}^\prime$  and find the momentum
\be
\pi^2_{L,\alpha}={R^2\over G^2} \left({R^{\prime2}\over L^2}-1+R^2 H_\alpha^2+{2G M_\alpha\over R} \right)={R^2\over G^2}\left({R^{\prime2}\over L^2} -A_{\alpha}\right)\,,\label{pireq}
\ee
or
\be
\pi_{L,\alpha}=\eta_{\pi} {R\over G}\sqrt{{R^{\prime2}\over L^2} -A_{\alpha}}\,,~~~\pi_{R,\alpha}={L\over R^\prime} \pi_{L,\alpha}^\prime\,,
\ee
where $\eta_{\pi}=\pm 1$, we defined the Hubble constants $H^2_\alpha=8\pi G\rho_\alpha/3$ and
\be
A_\alpha=1-R^2 H^2_\alpha -{2 G M_\alpha\over R}\,.
\ee
We now find the $2(N-1)$ matching conditions by integrating the Hamiltonian constraints (\ref{junction2a}) over the shell located at $\hat{r}_\alpha$,
\bea
\hat{p}_\alpha&=&-\int_{\hat{r}_\alpha-\epsilon}^{\hat{r}_\alpha+\epsilon} dr~L \pi_L^\prime+\dots=-\hat{L}_\alpha(\hat{\pi}_{L,\alpha,+}-\hat{\pi}_{L,\alpha,-})\,,\\
-{G\over \hat{R}_\alpha}\sqrt{{\hat{p}_\alpha^2}+m_\alpha^2\hat{L}_\alpha^2 }&=&{\hat{L}_\alpha\over \hat{R}_\alpha}\int_{\hat{r}_\alpha-\epsilon}^{\hat{r}_\alpha+\epsilon} dr~\left({R R^\prime\over L}\right)^\prime+\dots=\hat{R}^\prime_{\alpha,+}-\hat{R}^\prime_{\alpha,-}\,,
\eea
where the dots represent terms continuous in $r$ that do not contribute to the integral and quantities evaluated on the inside/outside of the shell are denoted by a hat and the index $+/-$, respectively. These are precisely the Israel junction conditions (\ref{junctionfulltext}) \cite{Israel:1966rt}.

\section{Classical Domain Wall Dynamics}\label{app2}
In this appendix we provide some supplemental material on the definition of  Gibbons-Hawking and Kruskal-Szekeres coordinates and the constraint equations governing the dynamics of bubble with a Schwarzschild interior and a de Sitter exterior, a true vacuum bubble.

\subsection{Gibbons-Hawking and Kruskal-Szekeres coordinates}

In order to illustrate the dynamics of domain walls that cross horizons we require coordinates that are smooth everywhere. Following \cite{Blau:1986cw}, we introduce Kruskal-Szekeres (KS) coordinates $(V_{\text{KS}},U_{\text{KS}},\theta_{\text{KS}},\phi_{\text{KS}})$ in domains with Schwarzschild metric and Gibbons-Hawking (GH) coordinates $(V_{\text{GH}},U_{\text{GH}},\theta_{\text{GH}},\phi_{\text{GH}})$ in domains with de Sitter metric. In the following we  drop the subscripts GH and KS since the distinction will be obvious from the context. The KS and GH coordinates are defined piecewise in four regions that, taken together, cover all of de Sitter and Schwarzschild space, respectively. These regions are defined as follows,
\bea
\text{I}&:~U>0\,,~|V|<|U|\,,~~~~~ \text{II}&:~V>0\,,~|U|<|V|\,,\nonumber\\ \text{III}&:~U<0\,,~~~~|V|<|U|\,,~\text{VI}&:~ V<0\,,~|U|<|V|\,.
\eea
The coordinates for Schwarzschild regions are defined as
\bea
U^\text{I}_{\text{KS}}=&\sqrt{{r\over 2GM}-1} e^{r/4GM}\cosh\left({t\over 4GM}\right)\nonumber\,,~~
V^\text{I}_{\text{KS}}=&\sqrt{{r\over 2GM}-1} e^{r/4GM}\sinh\left({t\over 4GM}\right)\nonumber\,,\\
U^\text{II}_{\text{KS}}=&\sqrt{1-{r\over 2GM}} e^{r/4GM}\sinh\left({t\over 4GM}\right)\nonumber\,,~~
V^\text{II}_{\text{KS}}=&\sqrt{1-{r\over 2GM}} e^{r/4GM}\cosh\left({t\over 4GM}\right)\nonumber\,,\\
U^\text{III}_{\text{KS}}=&-\sqrt{{r\over 2GM}-1} e^{r/4GM}\cosh\left({t\over 4GM}\right)\nonumber\,,~~
V^\text{III}_{\text{KS}}=&-\sqrt{{r\over 2GM}-1} e^{r/4GM}\sinh\left({t\over 4GM}\right)\nonumber\,,\\
U^{\text{IV}}_{\text{KS}}=&-\sqrt{1-{r\over 2GM}} e^{r/4GM}\sinh\left({t\over 4GM}\right)\nonumber\,,~~
V^{\text{IV}}_{\text{KS}}=&-\sqrt{1-{r\over 2GM}} e^{r/4GM}\cosh\left({t\over 4GM}\right)\,.
\eea
The Schwarzschild metric in Kruskal-Szekeres coordinates is given by
\be
ds^2={32 (G M)^3\over r}e^{-r/2GM} (-dV+dU^2)+r^2 d\Omega^2_2\,,
\ee
and $r$ is defined by
\be
U^2-V^2=\left({r\over 2 G M}-1\right)e^{r/2GM}\,.
\ee
Similarly, we have for the Gibbons-Hawking coordinates
\bea
U^\text{I}_{\text{GH}}=&\sqrt{1-H r\over 1+H r}\cosh\left(H t\right)\nonumber\,,~~
V^\text{I}_{\text{GH}}=&\sqrt{1-H r\over 1+H r}\sinh\left(H t\right)\nonumber\,,\\
U^\text{II}_{\text{GH}}=&\sqrt{H r-1\over 1+H r}\sinh\left(H t\right)\nonumber\,,~~
V^\text{II}_{\text{GH}}=&\sqrt{H r-1\over 1+H r}\cosh\left(H t\right)\nonumber\,,\\
U^\text{III}_{\text{GH}}=-&\sqrt{1-H r\over 1+H r}\cosh\left(H t\right)\nonumber\,,~~
V^\text{III}_{\text{GH}}=&-\sqrt{1-H r\over 1+H r}\sinh\left(H t\right)\nonumber\,,\\
U^{\text{IV}}_{\text{GH}}=&-\sqrt{H r-1\over 1+H r}\sinh\left(H t\right)\nonumber\,,~~
V^{\text{IV}}_{\text{GH}}=&-\sqrt{H r-1\over 1+H r}\cosh\left(H t\right)\,.
\eea
The de Sitter metric becomes 
\be
ds^2=\left({1+H r\over H}\right)^2(-dV^2+dU^2)+r^2 d\Omega^2_2\,,
\ee
where $r$ is given by
\be
U^2-V^2={1-H r\over 1+H r}\,.
\ee
The coordinate systems for both spacetimes are illustrated in Figure \ref{desitterschwarzschild}.

\subsection{Schwarzschild/de Sitter domain wall dynamcis}
\begin{figure}
  \centering
  \includegraphics[width=1\textwidth]{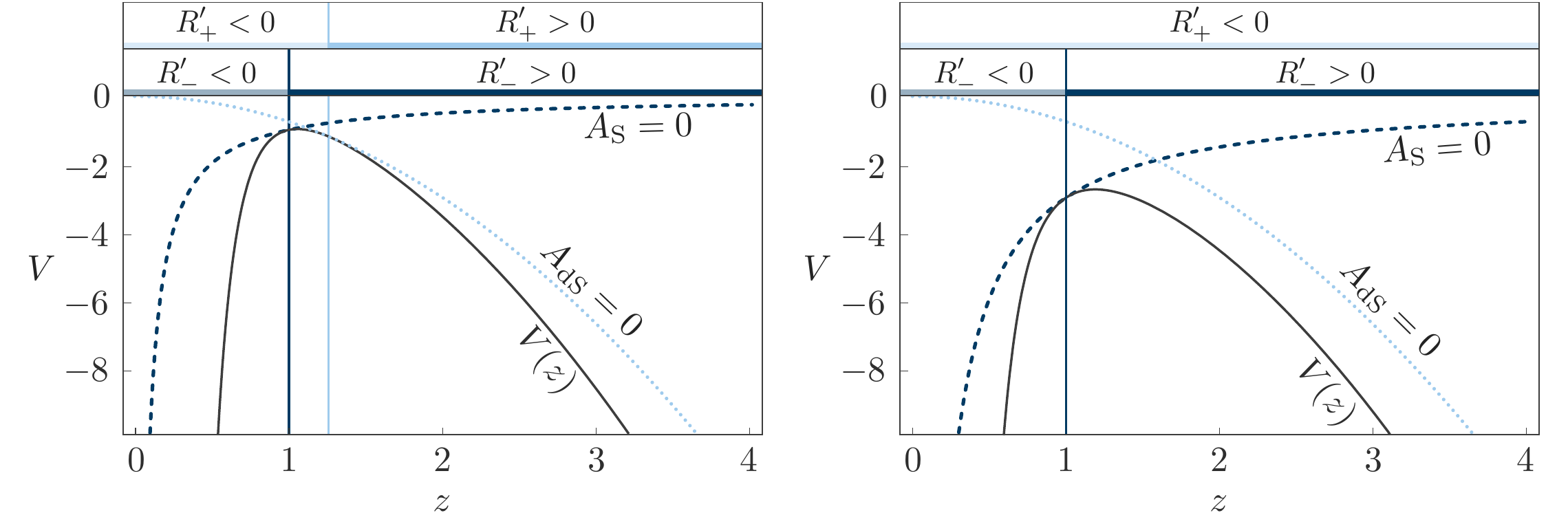}
  \caption{\small  Effective potential $V(z)$ as a function of rescaled radial coordinate $z$, for $\kappa<H$ (left) and $\kappa>H$ (right) for a Schwarzschild/de Sitter domain wall. The situation is equivalent to Figure \ref{figurefull2} after a coordinate change that reverses the inside and outside regions.}\label{figurefull}
\end{figure}

In this subsection we give the equations governing the dynamics of a spherical patch of Schwarzschild space that is separated by a thin wall from a de Sitter phase. While the dynamics are identical to the dynamics of a bubble of de Sitter phase within a Schwarzschild background, discussed in \S\ref{dSSdynamics} (by exchanging the definition of inside and outside), we provide the dynamics for the reverse situation for quick reference.

For the case of a Schwarzschild/de Sitter domain wall the rescaled radial coordinate $z$ is again given by (\ref{zdss}),
\be
z= \frac{{H}_+^2}{2 G M_{\cal A}} \hat{R}^3\,,~~{H}_+^2={4{H}_{}\over 2-\gamma}\,,~~\gamma={2}{\kappa^2-{H}_{}^2\over \kappa ^2+{H}^2}\,,~~ E=-\frac{4 \kappa ^2}{  (2G M_{S}) ^{2/3}{H}_+^{8/3}}\,,
\ee
but now the metric is given by (\ref{staticmetric}) with 
\bea\label{metricfactorsapp}
A_{\cal B}=1-{2 G M\over R}\,,~~~\text{for $r<\hat{r}$}\,,\nonumber\\
A_{\cal A}=1-{H}^2 R^2\,,~~~\text{for $r>\hat{r}$}\,,
\eea
Remember that $\gamma\approx -2$ when the domain wall tension in Planck units is small compared to the Hubble scale, 
\be
{\sigma^2\over \M^2}\ll {4\over 3} \rho\,.
\ee

\begin{figure}
  \centering
  \includegraphics[width=1\textwidth]{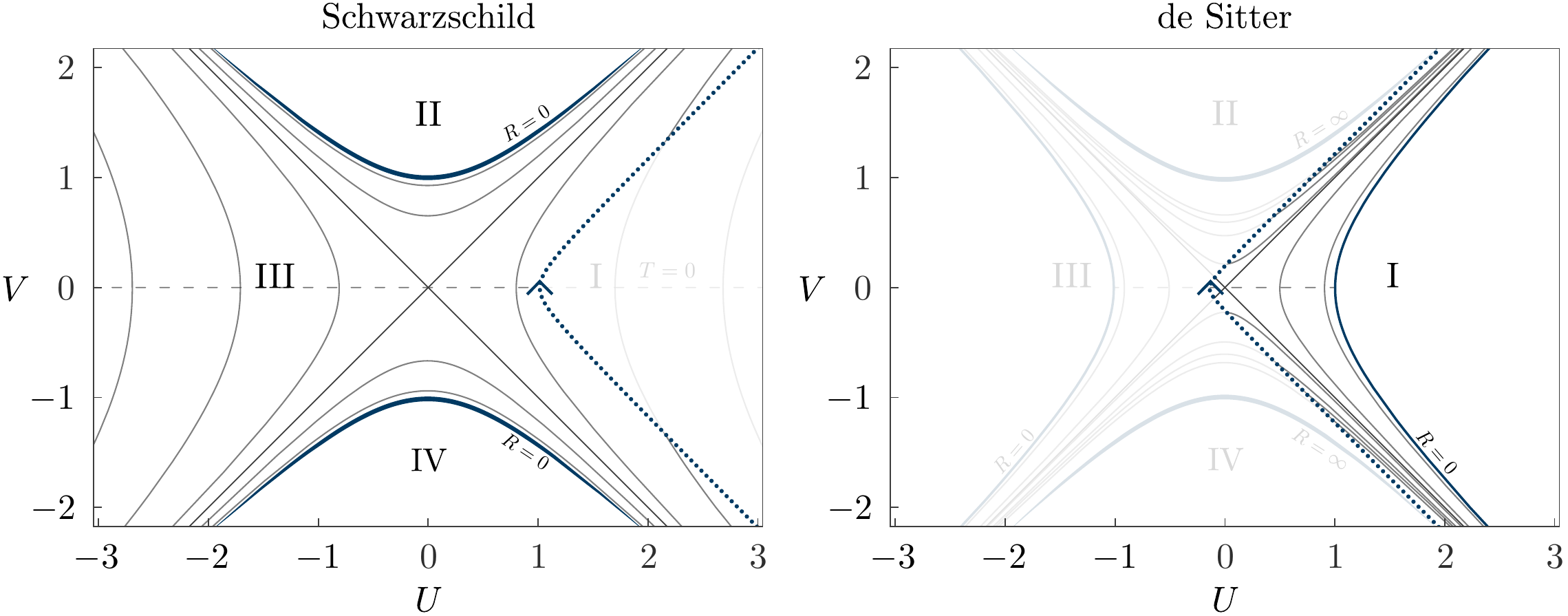}
  \caption{\small  Illustration of the unbound trajectory of a Schwarzschild/de Sitter domain wall that traverses the regions IV-III-II in the de Sitter diagram and is contained in region I of the Schwarzschild diagram. The spacetime diagram is not applicable in the faded region.}\label{Sdssols}
\end{figure}

We can write $\hat{R}^\prime$, and the position of the horizons in terms of the new radial coordinate $z$ as
\be
{\hat{R}^\prime_{-}}={z^3-1\over \sqrt{|E|} z^2}\,,~~{\hat{R}^\prime_{+}}=-{2+\gamma z^3\over 2\sqrt{|E|} z^2}\,, ~~z_{A_{\text{dS}}=0}={2+\gamma\over |E|}\,,~~z_{A_{\text{S}}=0}={\sqrt{|E|\over 1-\gamma^2/4}}\,.
\ee
As expected, these expressions are identical to (\ref{rprimeandstuff}) with the replacement of indices $-\,\rightarrow\,+$ and coordinates $r\rightarrow \text{const.}-r$. The full dynamics can be read off from the effective potential in Figure \ref{figurefull}. 
The effective potential has a maximum at
\be\label{zmax}
z_\text{max}^3={\gamma\over 4}+\sqrt{2+\left({\gamma\over 4}\right)^2}\,,~~V(z_\text{max})=-{3\over 2}\dfrac{ {\gamma} \sqrt{2+\left({\gamma\over 4}\right)^2}+{\gamma^2\over 4}+2}{\left(\sqrt{2+\left({\gamma\over 4}\right)^2}+{\gamma\over 4}\right)^{4/3}}\,.
\ee
An example of an unbound trajectory, as seen both in the Schwarzschild and de Sitter coordinates, is shown in Figure \ref{Sdssols}. 
All possible bound and unbound solutions are summarized in Table \ref{SdStrajectories}.
\begin{table*}
\begin{center}
\begin{tabular}{@{}p{2.5cm} @{}p{3.5cm}@{}p{3.5cm}  @{}p{4.8cm} }\\\toprule
Type& de Sitter & Schwarzschild  &Conditions\\\midrule
Bound&I&IV - III - II& $E<V_{\hat{R}^\prime_+=0}<V_\text{max}$ \\
Bound&I&IV - I - II& $V_{\hat{R}^\prime_+=0}<E<V_\text{max}$ \\
Unbound&I - II&IV - I& $V_\text{max}<E$\phantom{$V_{\hat{R}^\prime_+}$} \\
Unbound&IV - I - II&I& $V_{\hat{R}^\prime_-=0}<E<V_\text{max}$,\,~$\kappa<H$ \\
Unbound&IV - III - II&I& $E<V_{\hat{R}^\prime_-=0}<V_\text{max}$ \\
\bottomrule\\
\end{tabular}
\caption{\small \label{SdStrajectories} Summary of all possible trajectories of a Schwarzschild/de Sitter domain wall. We show the regions of the corresponding conformal diagrams that are traversed by the trajectories.}
\end{center}
\end{table*}

\section{Alternate Expression for the Tunneling Rate}\label{app3}
 Coleman and de Luccia computed the vacuum decay rate in the leading semiclassical approximation and including gravity in \cite{Coleman:1980aw}. However, their result for the tunneling rate, (3.12, 3.13) in \cite{Coleman:1980aw} only applies for domain wall extrinsic curvatures that do not change sign across the wall and transitions that do not contain any causally disconnected spacetimes. 
Their result was rewritten by Parke in a more compact form \cite{Parke:1982pm,Weinberg:2012pjx}.
 
We derived the transition probability using a Hamiltonian framework and obtained the tunneling exponent (\ref{tunnelingexp}). Our result is valid for arbitrary domain wall curvatures, but we wrote the expression in a more compact form. 
We now express the tunneling exponent in a way that closely resembles (3.13) of \cite{Coleman:1980aw},
\bea\label{alternatetunneling}
B
&=&\frac{3}{16 G^2} \Bigg(\frac{\text{sgn}(\hat{R}^\prime_{-})\left[1-\frac{8}{3} \pi  G {{\rho}}_{\cal B} \hat{R}^2\right]^{3/2}-1}{{{\rho}}_{\cal B}}- \frac{\text{sgn}(\hat{R}^\prime_{+})\left[1-\frac{8}{3} \pi  G {{\rho}}_{\cal A} \hat{R}^2\right]^{3/2}-1}{{{\rho}}_{\cal A}}\Bigg)\nonumber\\&&+2 \pi ^2 \hat{R}^3 \sigma+{3\over 8 G^2}\Bigg({N_{{\cal B},\,{\mathcal I}}-N_{{\cal B},\,{\mathcal F}}\over {{\rho}}_{\cal B}}+{N_{{\cal A},\,{\mathcal I}}-N_{{\cal A},\,{\mathcal F}}-\Theta(-\hat{R}^\prime_{+})\over {{\rho}}_{\cal A}}\Bigg)\,,
\eea
where $\hat{R}$ is given in (\ref{radiuss}) and the junction conditions (\ref{rpsimplegauge}) set the sign of the extrinsic curvature, 
\be
\text{sgn}(\hat{R}^\prime_{\pm})=\text{sgn}({{\rho}}_{{\cal A}}-{{\rho}}_{{\cal B}}\mp 6\pi G\sigma^2)\,.
\ee
In the absence of disconnected spacetimes the last term in (\ref{alternatetunneling}) vanishes.
For positive domain wall curvatures (\ref{alternatetunneling}) agrees with the result obtained by Coleman and de Luccia. In the limit of vanishing energy density inside the bubble, we find with (\ref{tunnelingexp}) for arbitrary domain wall curvatures
\be\label{blow}
B={27 \pi^2\sigma^4\over 2\rho_{\cal A}(\rho_{\cal A}+6 \pi G \sigma^2)^2}={B_0\over [1+(\hat{R}_0/2\Lambda)^2]^2}\,,
\ee
where $B_0=27\pi^2 \sigma^4/(2\rho_{\cal A}^3)$, $\hat{R}_0=3\sigma/\rho_{\cal A}$, and $\Lambda=(8\pi G \rho_{\cal A}/3)^{-1/2}$. The result (\ref{blow}) is identical to (3.16) --- but at large tensions disagrees with (3.13) --- of \cite{Coleman:1980aw}. Curiously, Coleman and de Luccia found with their (3.16) the correct answer for the both the large and small tension regimes, although they considered only the latter scenario and then made a sign ``error''. Similarly, the result given in \cite{Parke:1982pm} is identical to our general results (\ref{tunnelingexp}) and (\ref{alternatetunneling}) even though their calculation, again, only applies in the small tension regime.

\bibliographystyle{JHEP}
\bibliography{bubblerefs}
\end{document}